\documentclass[aps,prb,floatfix,twocolumn,superscriptaddress]{revtex4-2}

\usepackage{mathrsfs}
\usepackage{graphicx}
\usepackage[english]{babel}
\usepackage{amsmath}
\usepackage{amssymb}
\usepackage{xcolor}
\usepackage{cancel}
\usepackage[normalem]{ulem}

\usepackage{relsize}
\usepackage{CJK}
\usepackage{dsfont}
\usepackage{bm}

\newcommand{\me}{\mathrm{e}}
\newcommand{\mi}{\mathrm{i}}

\newcommand{\dif}{\mathrm{d}}

\newcommand\px{\mathrel{/\mkern-5mu/}}
\allowdisplaybreaks

\begin{document}

\title{Uhlmann quench and geometric dynamic quantum phase transition of mixed states}

\author{Jia-Chen Tang}
\affiliation{School of Physics, Southeast University, Jiulonghu Campus, Nanjing 211189, China}

\author{Xu-Yang Hou}
\affiliation{School of Physics, Southeast University, Jiulonghu Campus, Nanjing 211189, China}

\author{Zheng Zhou}
\affiliation{School of Physics, Southeast University, Jiulonghu Campus, Nanjing 211189, China}

\author{Hao Guo}
\email{guohao.ph@seu.edu.cn}
\affiliation{School of Physics, Southeast University, Jiulonghu Campus, Nanjing 211189, China}
\affiliation{Hefei National Laboratory, Hefei 230088, China}

\author{Chih-Chun Chien}
\affiliation{Department of physics, University of California, Merced, CA 95343, USA}

\begin{abstract}
Dynamic quantum phase transitions (DQPT) following quantum quenches exhibit singular behavior of the overlap between the initial and evolved states. Here we present a formalism to incorporate a geometric phase into quench dynamics of mixed quantum states, a process named the Uhlmann quench, based on the Uhlmann parallel transport. To overcome the incompatibility between the Uhlmann parallel-transport condition and Hamiltonian dynamics, we formulate the evolution of purification of the density matrix in a form which not only respects the dynamics according to the density matrix but also follows the Uhlmann parallel-transport condition to generate a geometric phase after a quantum quench. For cyclic processes exemplified by a spin-1/2 system, geometric DQPTs (GDQPTs) can emerge with both singular behavior in the dynamic analogue of the free energy and jumps of the geometric phase. Moreover, the Uhlmann phase reflecting the holonomy is generated at the end of each cycle. The Uhlmann quench thus paves the way for investigating the interplay between quantum dynamics and geometric processes in mixed states.
\end{abstract}

\maketitle

\section{Introduction}
Since the formulation of the Berry phase~\cite{Berry84}, geometric phase has become an important concept in physics. The underlying mechanism of geometric phase is closely related to many geometric and topological properties of quantum systems~\cite{PhysRevLett.51.2167,PhysRevA.36.3479,QGT10} and related fields of physics~\cite{Bohm03,Vanderbilt_book,Cohen19}. The geometric formalism has become an essential theoretical tool for studying and understanding topological quantum computation \cite{Kitaev_2001,RevModPhys.80.1083,AliceaNP11,SciPostPhys.3.3.021}, topological insulators and superconductors~\cite{KaneRMP,ZhangSCRMP,Bernevigbook,ChiuRMP,MooreN,KaneMele,KaneMele2,BernevigPRL,MoorePRB,FuLPRL}, and other emerging branches of physics~\cite{TKNN,Haldane,RevModPhys.82.1959}.
The question of extending the theory of geometric phase from pure states in the literature to mixed states naturally arises because mixed states are ubiquitous.
Some proposed generalizations include the Uhlmann phase~\cite{Uhlmann86,Uhlmann89,Uhlmann91} and the ``interferometric geometric phase"~\cite{PhysRevLett.85.2845}. Here, we focus on the Uhlmann phase since its formalism is analogous to that of the Berry phase because both can be derived from the concepts of parallel transport and principal bundles \cite{PhysRevB.91.165140,ourPRB20}. Importantly, the Berry phase and Uhlmann phase are genuine geometric objects because they reflect the holonomies of the underlying fiber bundles, respectively.

Analytical expressions of the Uhlmann phase of several exemplary topological models have been derived~\cite{PhysRevLett.112.130401}. Furthermore, temperature-induced topological quantum phase transitions (TQPTs) have been found in the exact solutions of those systems, where the Uhlmann phase exhibits quantized jumps.
This has sparked more research on the Uhlmann phase in other quantum systems \cite{PhysRevLett.113.076407,Uhlmann17,Uhlmann18,npj18,PhysRevLett.119.015702,PhysRevB.97.235141,PhysRevA.98.042316,OurPRB20b,Galindo21}. Meanwhile, the Uhlmann connection also plays a crucial role in understanding the local geometry of mixed quantum states~\cite{OurQGT23}.
Despite those achievements,
the formalism behind the Uhlmann phase has some disadvantages compared to that of the Berry phase.
For example, the Uhlmann bundle is topologically trivial~\cite{PhysRevB.91.165140}, forcing its characteristic classes to vanish. Moreover, the Uhlmann parallel-transport condition was found to be incompatible with Hamiltonian dynamics~\cite{ourPRB20}, and the Uhlmann phase requires transformations of both system and ancilla representing the environment~\cite{OurPRA21}. It is thus challenging to measure the Uhlmann phase in natural systems. However, simulations of the Uhlmann phase of a two-level system on quantum computers have shown promising results~\cite{npj18}.


While physical processes exhibiting TQPTs typically require additional geometric constraints, interesting dynamic processes due to sudden changes of parameters, known as quench dynamics, give rise to interesting time-dependent phenomena. Quantum quench has become a thriving area of study spanning nonequilibrium physics, quantum information, and condensed matter physics \cite{DQPT14,DQPT15,DQPTB2,DQPTPRB16,DQPTreview18}. One striking phenomenon emerging in this field is the dynamic quantum phase transition (DQPT), which exhibits nonanalytic behavior in the real-time dynamics of a quantum system after a quench of a parameter in time. At a DQPT, the Loschmidt amplitude of the overlap between the initial and evolved states vanishes, leading to divergence of the time-analogue of the free energy. The importance of DQPTs lies in their direct connections with observable behavior of quantum many-body systems in quench dynamics. Significantly, DQPTs have been realized in various quantum systems \cite{DQPTB41,DQPTB4,Zhang_2017,PhysRevApplied.11.044080,PhysRevLett.122.020501,PhysRevB.100.024310,PhysRevLett.124.250601,PhysRevLett.128.160504}.
As will be shown in the following, if a geometric phase is accumulated in the quench dynamics of mixed states by following certain geometric constraints, there may be jumps of the phase at each transition point, which will be called the geometric DQPT (GDQPT) to emphasize the dynamics is compatible with additional geometric constraints. 
Some topological aspects of DQPT for pure states have been studied in Ref.\cite{PhysRevB.93.085416,ZianiSR20}. 
To our knowledge, the physics of GDQPT for mixed states has not been systematically investigated in the literature. 

In the following, we present an intriguing combination of  the Uhlmann parallel-transport condition and quench dynamics due to a sudden change of the system Hamiltonian. Despite the incompatibility between Uhlmann's parallel transport and Hamiltonian dynamics~\cite{ourPRB20}, we successfully incorporate the Uhlmann parallel-transport condition into the time evolution process following a quench by introducing the dynamics of the purification of the density matrix, which respect the additional condition. Consequently, both dynamic and geometric phases are generated in a single physical process, which we refer to as the Uhlmann quench because of its compatibility with the Uhlmann parallel transport. We will show that the Uhlmann process can exhibit singular behavior corresponding to the DQPT of ordinary quench processes. Nevertheless, the geometric nature of the system changes at a GDQPT in the Uhlmann process as indicated by a jump of the geometric phase. Moreover, if the post-quench evolution is cyclic in time, the Uhlmann phase can be identified when the density matrix returns because of the associated holonomy. Therefore, this approach provides a protocol for generating and measuring the Uhlmann phase in dynamic systems. We present an explicit example to elucidate the details and subtleties of the Uhlmann process. Multiple GDQPTs of mixed states with jumps of the geometric phase at critical times in the example will illustrate interesting phenomena from quench dynamics with additional geometric constraints of mixed states.

The rest of the paper is as follows: In Section \ref{SecII}, we presents the theoretical framework, including purification of density matrices and the theory of the Uhlmann phase in Sec.~\ref{SecII.a}, the formalism of the quantum quench dynamics in Sec.~\ref{SecII.b}, and the formalism and protocol of the Uhlmann-quench in Sec.~\ref{SecII.c}. In Section \ref{SecIII}, we present an explicit example of a spin-$\frac{1}{2}$ system to illustrate the GDQPT from the Uhlmann quench and its features. Section \ref{SecIV} discusses some implications for experimentally realize and measure the Uhlmann quench.
Finally, Sec.~\ref{SecV} concludes our work.

\section{Theory of Uhlmann Quench}\label{SecII}

\subsection{Uhlmann process and TQPT}\label{SecII.a}
For simplicity, we will set $c=\hbar=k_B=1$ throughout the paper.
To generalize the formalism of geometric phase and quench dynamics to mixed quantum states, it is convenient to adopt a pure-state-like description through purification of density matrices \cite{Uhlmann86}. If a full-rank density matrix is diagonalized as $\rho=\sum_{n=1}^N\lambda_n|n\rangle\langle n|$, it can be decomposed as $\rho=WW^\dag$ by its purification $W=\sqrt{\rho}U=\sum_{n=1}^N\sqrt{\lambda_n}|n\rangle\langle n|U$. Here $U$ is an arbitrary unitary matrix and serves as the phase factor of the amplitude $W$. Moreover, there is a correspondence between $W$ and a state-vector representation called the purified state
\begin{align}\label{Wps}
|W\rangle=\sum_{n=1}^N\sqrt{\lambda_n}|n\rangle_s\otimes U^T|n\rangle_a.
\end{align}
Here the extra subscripts ``$s$'' and ``$a$'' are included to emphasize the system and ancilla states respectively. Thus, $U$ can be regarded as an operator acting on the auxiliary ancilla space. Moreover, $\rho$ represents the density matrix of the system states since
\begin{align}\label{rho1}
\rho=\text{Tr}_a(|W\rangle\langle W|),
\end{align}
where $\text{Tr}_a$ is the partial trace taken over the ancilla space.
The inner product between two purified states follows the Hilbert-Schmidt product $\langle W_1|W_2\rangle=\text{Tr}(W^\dag_1W_2)$.

A quantum system depending on a set of parameters $\mathbf{R}=\{R^1,\cdots R^k\}\in M$ is considered, where $M$ is the parameter manifold. A closed curve $C(t):=\mathbf{R}(t)$ ($0\le t\le \tau$) in $M$ introduces a process of $\rho$  and its purification as $t$ evolves. Here $t$ is a parameter of the process, and if $t$ is chose to be time, the process describes the evolution of the system. Explicitly,
\begin{align}\label{rt}
t\mapsto \rho(t),\quad t\mapsto W(t),\quad \rho(t)=W(t)W^\dag(t),
\end{align}
where $\rho(t)\equiv \rho(\mathbf{R}(t))$ and $W(t)\equiv W(\mathbf{R}(t))=\sqrt{\rho(\mathbf{R}(t))}U(\mathbf{R}(t))$. The curve $\gamma(t):=\rho(t)$ is closed since $\rho(\mathbf{R}(0))=\rho(\mathbf{R}(\tau))$. However, due to the arbitrariness of $U$, the curve $\tilde{\gamma}(t):=W(t)$ is open in general since $U(\mathbf{R}(\tau))$ is not necessarily equal to $U(\mathbf{R}(0))$. Their difference after parallel-transport through a cyclic process generates the Uhlmann phase.

The length of $\tilde{\gamma}$, given by $L(\tilde{\gamma})=\mathlarger{\int}_{\tilde{\gamma}}\sqrt{\langle\dot{W}|\dot{W}\rangle}\dif t$,
is minimized by the Uhlmann parallel-transport condition
\begin{align}\label{pc}
\dot{ W}W^\dag=W\dot{W}^\dag.
\end{align}
In this case, $\tilde{\gamma}$ represents the horizontal lift of $\gamma$, and $W$ (or its phase factor $U$) is said to be parallel-transported along $\tilde{\gamma}$.
We refer to a physical process satisfying Eq.~(\ref{pc}) as the Uhlmann process.
Let $X$ be the tangent vector of $\gamma$, then the Uhlmann parallel-transport condition is equivalent to \cite{ourPRB20}
\begin{align}\label{ptU}
\nabla_XU(t)=\frac{\dif U(t)}{\dif t}+A_U(X)U(t)=0,
\end{align}
where
\begin{align}\label{AU}
A_U=-\sum_{n,m}|n\rangle\frac{\langle n|[\dif\sqrt{\rho},\sqrt{\rho}]|m\rangle}{\lambda_n+\lambda_m}\langle m|
\end{align}
is the Uhlmann connection. The theory of Uhlmann phase can be formulated using the language of a U($N$) principal bundle briefly outlined in Appendix \ref{appa00}. On the base of the fiber bundle, every continuous curve $\gamma(t)$ has a unique horizontal lift $\tilde{\gamma}(t)$ with respect to the connection $A_U$, which is independent of the choice of the ancilla, as Eq.~\eqref{AU} suggests.

Eq.~(\ref{ptU}) indicates that $A_U$ is a non-Abelian gauge connection and defines how the phase factor $U(t)$ is parallel-transported.
At the end of the parallel transport, the solution to Eq.~(\ref{ptU}) gives the difference between the initial and final phase factors, known as the Uhlmann holonomy
\begin{align}\label{U}
U(\tau)=\mathcal{P}\me^{-\oint_\gamma A_U }U(0).
\end{align}
Here $\mathcal{P}$ is the path-ordering operator.
The difference between the initial and final purifications is characterized by the Loschmidt (or transition) amplitude
\begin{align}
\mathcal{G}_U=\langle W(0)|W(\tau)\rangle
=\text{Tr}\left[\rho(0)\mathcal{P}\me^{-\oint_\gamma A_U }\right],
\end{align}
whose argument is the Uhlmann phase
\begin{align}\label{thetaUb}
\theta_U=\arg\mathcal{G}_U=\arg\text{Tr}\left[\rho(0)\mathcal{P}\me^{-\oint_\gamma A_U }\right].
\end{align}
According to Eq.~(\ref{U}), the geometric effect carried by the Uhlmann phase is introduced by $U$, which acts on the ancilla state.
When the evolution of a quantum system crosses the zeros of $\mathcal{G}_U$, the Uhlmann phase exhibits a discrete jump, signaling a TQPT because the Uhlmann holonomy has changed. Some examples of the TQPT of the Uhlmann process have been shown in Refs.~\cite{OurPRA21,Galindo21}.

\subsection{Quantum quench dynamics and DQPT}\label{SecII.b}
In a typical scenario of quantum quench dynamics, a quantum system is initially prepared in the state $|\psi(0)\rangle$. At $t=0^+$, the system undergoes a sudden quench due to a change in the Hamiltonian $H$, and $|\psi(0)\rangle$ is in general not an eigenstate of $H$. The key object measuring the deviation of the time-evolved state from the initial state is the Loschmidt (or return) amplitude
\begin{align}\label{LA}
\mathcal{G}(t)=\langle\psi(0)|\psi(t)\rangle=\langle\psi(0)|\me^{-\mi Ht}|\psi(0)\rangle.
\end{align}
In non-equilibrium physics, $|\mathcal{G}(t)|^2$ plays a role analogous to the partition function in thermodynamics~\cite{DQPTreview18}. Accordingly, we can introduce the counterpart of the free-energy in the theory of DQPT as
 \begin{align}\label{RF}
r(t)=-\lim_{N\rightarrow \infty}\frac{1}{N}\ln|\mathcal{G}(t)|^2,
\end{align}
where $N$ is the overall degrees of freedom. The critical time $t^*_n$ at which $r(t)$ exhibits singular behavior signify a dynamic quantum phase transition (DQPT) because $t^*_n$ is a zero of $\mathcal{G}(t)$.

The theory of DQPT can be readily extended to mixed states via purification. We consider a quantum system initially in the state described by $\rho(0)=W(0)W^\dag(0)$.
After diagonalizing the initial density matrix, $\rho(0)=\sum_n\lambda_n|n\rangle\langle n|$ and
 \begin{align}\label{W0}W(0)=\sqrt{\rho(0)}U(0)=\sum_n\sqrt{\lambda_n}|n\rangle\langle n|U(0).\end{align}
Following a quench governed by $H$, the density matrix evolves according to the Heisenberg equation $\rho(t)=\me^{-\mi Ht}\rho(0)\me^{\mi Ht}$.
There are some subtleties when decomposing $\rho(t)$. If the following decomposition is imposed,
 \begin{align}\label{decom1}
\rho(t)=W(t)W^\dag(t)=\me^{-\mi Ht}W(0)W^\dag(0)\me^{\mi Ht},
\end{align}
a simple choice of the purification evolution is
 \begin{align}\label{Wt0}
W(t)=\me^{-\mi Ht}W(0).
\end{align}
With this choice, the purification $W$ formally satisfies the following Schr\"odinger-like equation
 \begin{align}\label{Wt2}
\mi\dot{W}=H W.
\end{align}
By generalizing Eq.~(\ref{LA}), the Loschmidt amplitude in the case of mixed states is given by
 \begin{align}\label{Gtd}
 \mathcal{G}(t)
 &=\langle W(0)|W(t)\rangle=\text{Tr}\left[W^\dag(0)W(t)\right]\notag\\
 &=\text{Tr}\left[\rho(0)U(t)U^\dag(0)\right]=\text{Tr}\left[\rho(0)\me^{-\mi Ht}\right].
\end{align}
When the expression is compared to Eq.~(\ref{thetaUb}), the argument of $ \mathcal{G}(t)$ represents the dynamic phase accumulated during the evolution following the quench.

We refer to the choice of Eq.~\eqref{Wt0} and its dynamic process described by Eq.~(\ref{Wt2}) as Hamiltonian dynamics, in which the density matrix and its purification evolve according to the Heisenberg equation and Schr\"odinger-like equation, respectively.
However, using $\sqrt{\rho(t)}=\me^{-\mi Ht}\sqrt{\rho(0)}\me^{\mi Ht}$ and $W(t)=\sqrt{\rho(t)}U(t)=\me^{-\mi Ht}\sqrt{\rho(0)}\me^{\mi Ht}U(t)$, the phase factor evolves according to
 \begin{align}\label{Ut0}
U(t)=\me^{-\mi Ht}U(0).
\end{align}
Apparently, this expression differs from the evolution of the phase from the Uhlmann process described by Eq.~(\ref{U}). In fact, it can be shown that Hamiltonian dynamics is incompatible with the Uhlmann process \cite{ourPRB20}. A sketch of the mathematical proof is outlined in Appendix \ref{appa0}.

\subsection{Beyond Hamiltonian dynamics}
Our goal is to formulate a time-evolving post-quench process that is consistent with the Uhlmann parallel transport. However, such an implementation appears to be hindered by the inherent incompatibility between Hamiltonian dynamics and Uhlmann processes~\cite{ourPRB20} if $W$ follows the Schr\"odinger-like equation~\eqref{Wt2}.
Upon tracing its origin, we discover that the incompatibility stems from the choice of Eq.~(\ref{Wt0}).
In order to address this challenge, we attempt to decompose $\rho(t)$ beyond the conventional form (\ref{decom1}) according to
 \begin{align}\label{decom2}
\rho(t)&=W(t)W^\dag(t)\notag\\&=\me^{-\mi Ht}W(0)\mathcal{U}(t)\mathcal{U}^\dag(t)W^\dag(0)\me^{\mi Ht},
\end{align}
where a unitary operator $\mathcal{U}$ has been introduced. The decomposition leads to
 \begin{align}\label{Wt}
W(t)=\me^{-\mi Ht}W(0)\mathcal{U}(t).
\end{align}
Here, the evolution of $W$ follows
 \begin{align}\label{Wt3}
\mi\dot{W}=H W+\mi W \mathcal{U}^\dag \dot{\mathcal{U}}.
\end{align}
Importantly, the density matrix still evolves according to $\rho(t)=\me^{-\mi Ht}\rho(0)\me^{\mi Ht}$.

In terms of purified states, Eq.~(\ref{Wt}) reads
  \begin{align}\label{Wt3b}
|W(t)\rangle=\sum_n\sqrt{\lambda_n}\me^{-\mi Ht}|n\rangle_s\otimes \left[U(0)\mathcal{U}(t)\right]^T|n\rangle_a,
\end{align}
where Eqs.~(\ref{Wps}) and (\ref{W0}) have been applied. Hence, it becomes evident that non-dynamic effects can be introduced via $\mathcal{U}(t)$ while $\rho(t)$, the density matrix of the system, still evolves according to the Heisenberg equation. This opens up the possibility of incorporating the Uhlmann parallel-transport condition in Hamiltonian dynamics. Moreover, by utilizing Eq.~(\ref{Wt}) and $W(t)=\sqrt{\rho(t)}U(t)=\me^{-\mi Ht}\sqrt{\rho(0)}\me^{\mi Ht}U(t)$, we derive the following generic relation between $\mathcal{U}(t)$ and the phase factor:
 \begin{align}\label{UUt2}
U(t)=\me^{-\mi Ht}U(0)\mathcal{U}(t).
\end{align}
Thus, the phase factor also acquires additional effects carried by $\mathcal{U}(t)$ beyond dynamic evolution and introduce non-dynamic phase according to Eq.~(\ref{Gtd}).

\subsection{Formalism of Uhlmann quench}\label{SecII.c}
Here we explicitly show the construction of $\mathcal{U}(t)$ to incorporate Uhlmann's parallel-transport condition in Hamiltonian dynamics.
In a non-trivial Uhlmann process, $A_U(X)\neq 0$ in general. Hence, Eq.~(\ref{AU}) implies that $\dot{\sqrt{\rho}}\neq 0$, or equivalently, $\dot{\rho}\neq 0$. In order for the Uhlmann process to hold, it is then required that $\mi\dot{\rho}=[H,\rho]\neq 0$, which implies that $H$ is time-dependent. For simplicity, we consider a sudden quench of the Hamiltonian which changes abruptly at $t=0^+$ and remains unchanged afterward. Clearly, this satisfies $[\rho(0^+),H]\neq 0$. Furthermore, if the Uhlmann parallel-transport condition is respected by the quench dynamics, we refer to such a process as a Uhlmann quench. To generate the Uhlmann phase, which is related to the holonomy of the underlying Uhlmann bundle~\cite{ourPRB20}, we will require the dynamics following a Uhlmann quench to be cyclic. In other words, there exists a period $\tau>0$ such that $\rho(\tau)=\rho(0)$.

To explicit construct an Uhlmann quench, we substitute Eq.~(\ref{Wt}) into the Uhlmann parallel-transport condition of Eq.~(\ref{pc}), which yields
 \begin{align}
2\mi W^\dag H W=W^\dag W \mathcal{U}^\dag \dot{\mathcal{U}}+\mathcal{U}^\dag \dot{\mathcal{U}} W^\dag W.
\end{align}
For simplicity, we let the initial phase factor be $U(0)=1$.
After using $W(0)=\sqrt{\rho(0)}$ and $W^\dag W=\mathcal{U}^\dag \rho(0) \mathcal{U}$, the Uhlmann parallel-transport condition is then equivalent to
\begin{align}\label{tmp1}
 2\mi H=\sqrt{\rho(0)}\dot{\mathcal{U}}\mathcal{U}^\dag\sqrt{\rho(0)}^{-1}+\sqrt{\rho(0)}^{-1}\dot{\mathcal{U}}\mathcal{U}^\dag\sqrt{\rho(0)}.
\end{align}
By taking derivative with respect to $t$ in Eq.~(\ref{UUt2}), we get
 \begin{align}\label{UU}
\dot{\mathcal{U}}\mathcal{U}^\dag =\mi H +\me^{\mi Ht}\dot{U}U^\dag \me^{-\mi Ht}.
\end{align}
Using Eq.~(\ref{ptU}), the condition (\ref{tmp1}) further reduces to the following identity:
 \begin{align}\label{tmp2d}
2 \sqrt{\rho(0)} H \sqrt{\rho(0)}=\{\rho(0), H+\me^{\mi Ht}\mi A_U(X) \me^{-\mi Ht}\}.
\end{align}
Here $A_U$ is given by Eq.~(\ref{AU}). The proof is outlined in Appendix \ref{appa}.

During an Uhlmann process, the phase factor evolves as
 \begin{align}\label{Ut0cc}
U(t)=\mathcal{T}\me^{-\int_0^t\dif t'A_U(X(t'))\dif t'}U(0),
\end{align}
which agrees with Eq.~(\ref{U}).
Therefore, the Uhlmann parallel-transport condition holds during the quench dynamics as long as the purification evolves as
 \begin{align}\label{Wtf}
 W(t)&=\me^{-\mi Ht}W(0)\me^{\mi Ht}U(t)\notag\\&=\me^{-\mi Ht}W(0)\me^{\mi Ht}\mathcal{T}\me^{-\int_0^t\dif t'A_U(X(t'))\dif t'}
\end{align}
according to Eqs.~(\ref{Wt}) and (\ref{UUt2}). In this scenario, the unitary evolution $\mathcal{U}(t)$ is give by
 \begin{align}\label{Utf}
\mathcal{U}(t)=\me^{\mi Ht}U(t)=\me^{\mi Ht}\mathcal{T}\me^{-\int_0^t\dif t'A_U(X(t'))\dif t'}.
\end{align}
Therefore, an Uhlmann quench encompasses both Hamiltonian dynamics and Uhlmann process since the Uhlmann parallel-transport condition is followed.

In an Uhlmann quench, the Loschmidt amplitude~\eqref{Gtd} becomes
\begin{align}\label{GtU}
 \mathcal{G}(t)
 &=\text{Tr}\left[\sqrt{\rho(0)}\me^{-\mi Ht}\sqrt{\rho(0)}\mathcal{U}(t)\right]\notag\\
 &=\text{Tr}\left[\sqrt{\rho(0)}\me^{-\mi Ht}\sqrt{\rho(0)}\me^{\mi Ht}\mathcal{T}\me^{-\int_0^t\dif t'A_U(X(t'))\dif t'}\right].
\end{align}
Since both the dynamic and geometric effects are generated during an Uhlmann-quench, the argument of the Loschmidt amplitude gives the total phase $\theta_\text{tot}$, which includes not only the dynamic phase but also the geometric phase.
To isolate the geometric phase, we need to separate the dynamic part from $\theta_\text{tot}$. We note that Eq.~(\ref{Wtf}) represents a combined evolution. For simplicity, $U(t)=1$ is set, implying $\mathcal{U}(t)=\me^{\mi Ht}$ and $A_U=0$. This ensures that the corresponding post-quench evolution carries only a trivial Uhlmann process and is thereby purely dynamic. Thus,
 \begin{align}\label{WtfD}
 W_\text{D}(t)=\me^{-\mi Ht}W(0)\me^{\mi Ht}.
\end{align}
Taking the argument of Eq.~(\ref{GtU}) for this particular case, the dynamic phase is given by
 \begin{align}\label{Gtd1}
 \theta_\text{D}(t)&=\arg\langle W_\text{D}(0)|W_\text{D}(t)\rangle\notag\\
&=\arg\text{Tr}\left[\sqrt{\rho(0)}\me^{-\mi Ht}\sqrt{\rho(0)}\me^{\mi Ht}\right]\notag\\
 &=\arg\text{Tr}\left[\sqrt{\rho(0)}\sqrt{\rho(t)}\right].
\end{align}
Interestingly, since $\sqrt{\rho(0)}$ and $\sqrt{\rho(t)}$ are both Hermitian and positive-definite, $\theta_\text{D}(t)=0$. This is understandable because in the purified-state representation of Eq.~(\ref{WtfD}), $|W_\text{D}(t)\rangle=\sum_n\sqrt{\lambda_n}\me^{-\mi Ht}|n\rangle_s\otimes \left[\me^{\mi Ht}\right]^T|n\rangle_a$, so the contributions from the system and ancilla states cancel each other.
Consequently, for a generic $U(t)$, the geometric phase is obtained by eliminating the dynamic phase from $\theta_\text{tot}$. Explicitly,
 \begin{align}\label{UPcp}
 \theta_\text{G}(t)&=\theta_\text{tot}(t)- \theta_\text{D}(t)=\theta_\text{tot}(t).
\end{align}
At the zeros $t^*_n$ of $\mathcal{G}(t)$, the value of $\theta_\text{G}(t)$ experiences a discrete jump, signaling a dynamic quantum phase transition. Since  $\theta_\text{G}(t)$ represents the geometric phase in an Uhlmann quench, we refer to the transition with a change of the geometric phase as a geometric dynamic quantum phase transition (GDQPT).

The Uhlmann phase $\theta_U$ is generated only in a cyclic process because of the definition of the holonomy on a fiber bundle~\cite{Nakahara,ourPRB20}. Assuming the period of $\rho(t)$ is $\tau$, so $\rho(\tau)=\rho(0)$, then $\theta_\text{G}(t)$ is the Uhlmann phase at $t=n\tau$ with $n$ being a positive integer.
We remark that there is no guarantee that $t^*_n= n\tau$ (or equivalently, $\tau\neq t^*_{n+1}-t^*_n$ in general), hence the value of $\theta_U$ may not experience a sudden change at the GDQPT transition point $t^*_n$. Furthermore, the Uhlmann phase is a topological invariant associated with the Uhlmann holonomy and differs from the dynamic topological order parameter~\cite{PhysRevB.93.085416} for pure quantum states. The latter is defined via the winding number evaluated by using the associated geometric phase and changes abruptly at DQPT points. Finally, we point out that periodic dynamics with period $\tau$ may exist in certain systems because they only require $[\me^{\mi H\tau},\rho(0)]=0$ to ensure $\rho(\tau)=\rho(0)$. This condition may be achieved in systems with rotational symmetries. We remark that the condition does not necessarily imply $[H,\rho(0)]=0$.

\section{Example: Spin-$\frac{1}{2}$ systems}\label{SecIII}
For a better understanding of the Uhlmann quench, we present an explicit example of a quantum spin-$\frac{1}{2}$ paramagnet in the presence of a magnetic field $\mathbf{B}$. The Hamiltonian is $H=\omega_0\hat{\mathbf{B}}\cdot \mathbf{J}$, where $\omega_0$ is the Larmor frequency, $\hat{\mathbf{B}}=\mathbf{B}/|\mathbf{B}|$, and $\mathbf{J}=\frac{1}{2}\boldsymbol{\sigma}$. We assume that initially $\mathbf{B}\px \hat{e}_z$, so $H_0=\frac{\omega_0}{2}\sigma_z$. Therefore, the eigenvalues and eigenstates are
\begin{align}
E_-=-\frac{\omega_0}{2}, \quad |-\rangle=\begin{pmatrix}0\\1\end{pmatrix};\quad E_+=\frac{\omega_0}{2}, \quad |+\rangle=\begin{pmatrix}1\\0\end{pmatrix}.
\end{align}
The initial density matrix is
\begin{align}\label{rho0}
\rho(0)=\frac{1}{Z}\me^{-\beta H_0}=f(-\omega_0)|-\rangle\langle-|+f(\omega_0)|+\rangle\langle+|.
\end{align}
where $\beta=1/T$ is the inverse temperature, and $f(x)=1/(e^{\beta x}+1)$ is the Fermi distribution function.

The quench is introduced by a sudden change of the direction of the magnetic field at $t=0^+$ to $\mathbf{B}=B(\sin\theta\cos\phi, \sin\theta\sin\phi,\cos\theta)^T$. After the quench, the system is governed by the Hamiltonian
\begin{align}\label{sjHa}
H=\frac{\omega_0}{2}\begin{pmatrix}\cos\theta &\sin\theta\me^{-\mi\phi}\\\sin\theta\me^{\mi\phi}&-\cos\theta\end{pmatrix},
\end{align}
and the time-evolution operator is
\begin{align}
\me^{-\mi Ht}=\cos\left(\frac{\omega_0 t}{2}\right)-\mi\sin\left(\frac{\omega_0 t}{2}\right)\hat{\mathbf{B}}\cdot \boldsymbol{\sigma}.
\end{align}
Accordingly, the post-quench density matrix evolves as $\rho(t)=\me^{-\mi Ht}\rho(0)\me^{\mi Ht}$. Using Eqs.~(\ref{dAU}), (\ref{rho0}), and
\begin{align}
H=\me^{-\mi H t}H\me^{\mi H t}=\sum_{n,m=\pm}\me^{-\mi H t}|n\rangle\langle n|H|m\rangle\langle m|\me^{\mi H t},\notag
\end{align} the Uhlmann connection becomes
\begin{align}\label{dAU4}
\mi A_U(X(t))=\me^{-\mi H t}\chi\tilde{H}\me^{\mi H t}.
\end{align}
Here $\chi=\frac{2\me^{\frac{\beta\omega_0}{2}}}{\me^{\beta\omega_0}+1}-1$ and $\tilde{H}=|-\rangle\langle -|H|+\rangle\langle +|+|+\rangle\langle +|H|-\rangle\langle -|$. We denote the Uhlmann holonomy by $g(t)=\mathcal{T}\me^{-\mathlarger{\int}_0^t\dif t' A_U(X(t'))}$, which satisfies $g(0)=1$ and
\begin{align}\label{dAU5}
\frac{\dif g(t)}{\dif t}=- A_U(X(t))g(t)=\mi\me^{-\mi H t}\chi\tilde{H}\me^{\mi H t}g(t).
\end{align}
Although $g(t)$ involves the time-ordering operator $\mathcal{T}$, its analytic expression can still be obtained by noticing that both $H$ and $\tilde{H}$ are independent of $t$. Let $g'(t)=\me^{\mi H t}g(t)$, then $g'(0)=1$ and
\begin{align}
\frac{\dif g'(t)}{\dif t}&=\mi H \me^{\mi H t}g(t)+\me^{\mi H t}\frac{\dif g(t)}{\dif t}\notag\\
&=\mi( H+\chi\tilde{H})g'(t).
\end{align}
Solving this equation, we get
\begin{align}
g'(t)=\me^{\mi( H+\chi\tilde{H})t}g'(0)=\me^{\mi( H+\chi\tilde{H})t},
\end{align}
which further implies
\begin{align}\label{gt}
g(t)=\me^{-\mi H t}\me^{\mi( H+\chi\tilde{H})t}.
\end{align}

Using Eq.~(\ref{GtU}), a straightforward evaluation shows that the Loschmidt amplitude is
\begin{align}\label{tmp5b}
&\mathcal{G}(T,t)=\text{Tr}\left[\sqrt{\rho(0)}\me^{-\mi H t}\sqrt{\rho(0)}\me^{\mi(H+\chi\tilde{H})t}\right]\notag\\
=&\frac{4\sin^{2}{\theta}\sin(\frac{\omega_{0}t}{2}
)\sin(\frac{\omega_{0}t}{2}
B)}{(\me^{\frac{\beta \omega_{0}}{2}}+\me^{\frac{-\beta \omega_{0}}{2}})^{2}B}\notag\\+&\cos\left(\frac{\omega_{0}t}{2}
B\right)\left(\frac{A(t)}{1+\me^{-\beta \omega_{0}}}+\frac{A^{*}(t)}{1+\me^{\beta \omega_{0}}}\right)\notag\\
+&\frac{\mi\cos{\theta}\sin(\frac{\omega_{0}t}{2}B
)}{B}\left(\frac{A^{*}(t)}{1+\me^{\beta \omega_{0}}}-\frac{A(t)}{1+\me^{-\beta \omega_{0}}}\right),
\end{align}
where
\begin{align}
A(t)&=\cos\left(\frac{\omega_{0}t}{2}\right)+\mi\cos{\theta}\sin\left(\frac{\omega_{0}t}{2}\right),\notag\\B&=\sqrt{\cos^{2}\theta+(\chi+1)^2\sin^{2}\theta}.
\end{align}
Apparently, this result is independent of the longitude $\phi$, which is reasonable due to the rotational symmetry of the system.
To simplify the discussions, we examine a particular example with $\theta=\frac{\pi}{2}$, indicating that the end point of the quenched magnetic field lies on the equator. In this situation, Eq.~(\ref{tmp5b}) simplifies to:
\begin{align}\label{tmp5c}
\mathcal{G}(T,t)&=\cos\left[\frac{\omega_{0}}{2}(\chi+1) t\right]\cos\left(\frac{\omega_{0}t}{2}\right)\notag\\&+\frac{2\sin\left[\frac{\omega_{0}}{2}(\chi+1) t\right]}{\me^{\frac{\beta\omega_{0}}{2}}+\me^{\frac{-\beta\omega_{0}}{2}}}\sin\left(\frac{\omega_{0}t}{2}\right).
\end{align}

In practice, determining the zeros $t^*_n$ of $\mathcal{G}(T,t)$ typically requires numerical methods, as $\mathcal{G}(T,t)=0$ represents a transcendental equation with respect to $t$. Furthermore, the evolution of $\rho(t)$ following the Uhlmann quench is also periodic, as will be explained below. Without loss of generality, we opt for $\phi=0$ due to the symmetry of the quenched system. Hence, the density matrix evolves according to
\begin{align}\label{rhot}
\rho(t)=\me^{-\mi Ht}\rho(0)\me^{\mi Ht}=\frac{1}{Z}\me^{-\beta H(t)}.
\end{align}
Here
\begin{align}\label{Ht}
H(t)&=\me^{-\mi Ht} H_0\me^{\mi Ht}= \frac{\omega_0}{2}\me^{-\mi \frac{\omega_0 t}{2}\sigma_x}\sigma_z\me^{\mi \frac{\omega_0 t}{2}\sigma_x} \notag\\
&= \frac{\omega_0}{2}\Big[\cos(\omega_0t )\sigma_z-\sin(\omega_0t)\sigma_y\Big],
 \end{align}
which is equivalent to a rotation of $H_0=\frac{\omega_0}{2}\sigma_z$ about the $x$-axis. Therefore, $H(t)$ is cyclic with a period  $\tau=\frac{2\pi}{\omega_0}$, and so is $\rho(t)$. Based on the previous discussions, at $t=n\tau$, $\rho(n\tau)=\rho(0)$ and $\theta_\text{G}(T,n\tau)=\arg\mathcal{G}(T,n\tau)$ is the Uhlmann phase since the cyclic condition is satisfied.

\begin{figure*}[th]
\centering
\includegraphics[width=6.0in]{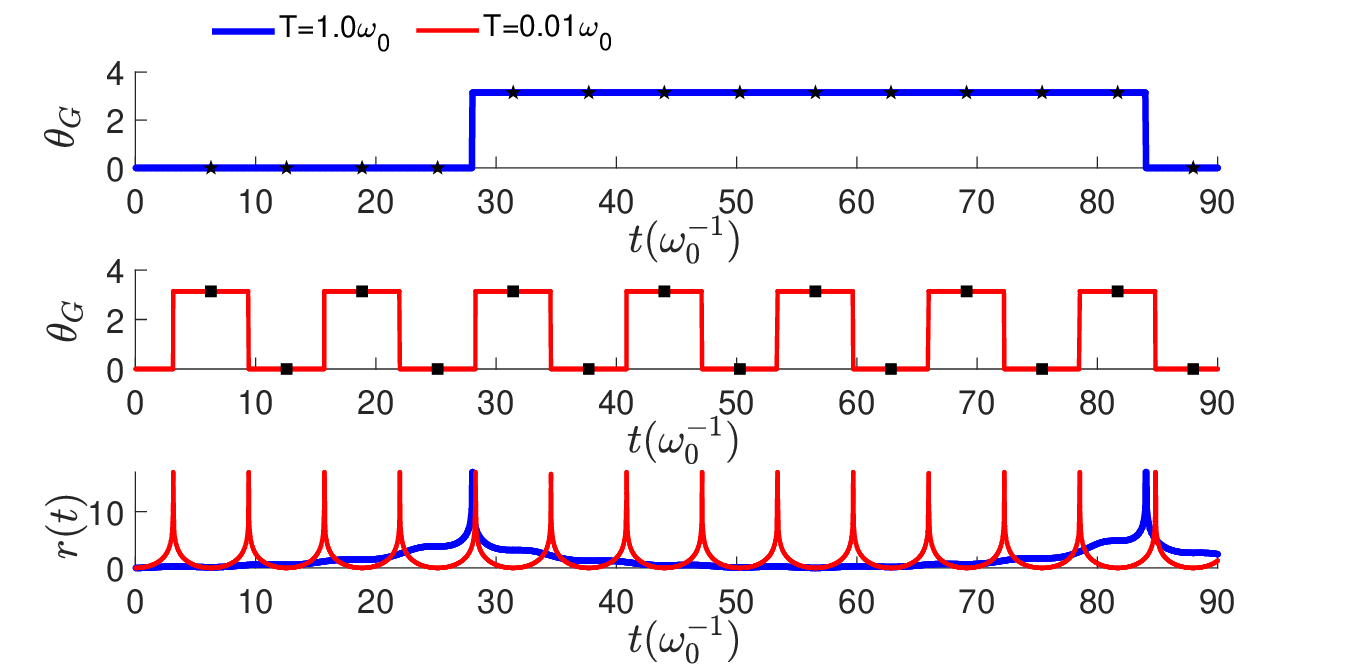}
\caption{ The geometric phase as a function of $t$ respectively at relatively high ($T=1.0\omega_0$, blue line in the top panel) and low ($T=0.01\omega_0$, red line in the middle panel) temperatures. The stars and squares label where the geometric phase is equal to the Uhlmann phase when the cyclic condition is satisfied.
(Bottom panel) The corresponding behavior of the rate function. The time $t$ is in units of $\omega_0^{-1}$.
}
\label{Fig1}
\end{figure*}

To visualize our results, We plot the Uhlmann phase $\theta_\text{G}$ and the rate function $r(t)$
as functions of $t$ in Figure~\ref{Fig1}. To be consistent with lattice models, we consider an effective lattice with $N$ sites. Each site hosts a spin-$1/2$ paramagnet with no interaction between different sites. Therefore, the total Loschmidt amplitude is $\mathcal{G}_{N}(T,t)=\mathcal{G}(T,t)^N$ due to the tensor product structure, and $r(t)=-\ln|\mathcal{G}(T,t)|^2$ with $\mathcal{G}(T,t)$ given by Eq.~\eqref{tmp5c}. The red and blue solid lines in Fig.~\ref{Fig1} represent the low- ($T=0.01\omega_0$) and relative high- ($T=1.0\omega_0$) temperature scenarios, respectively.

In the high-temperature limit, $\beta\rightarrow 0$ and $\chi\rightarrow 0$, so either Eq.~(\ref{tmp5b}) or (\ref{tmp5c}) yields $\mathcal{G}(T,t)=1$. This is reasonable since $\lim_{T\rightarrow+\infty} \rho=\frac{1}{2}1_2$ is proportional to the identity matrix and carries no physical information.
Hence, we choose a moderately high temperature $T=1.0\omega_0$ such that the behavior of system is nontrivial. We plot $\theta_\text{G}$ as a function of $t$ in the top panel and the corresponding $r(t)$ in the bottom panel by the blue solid lines. Specifically, the rate function $r(t)$ becomes singular at each crossing of $t^*_n$, and the value of the geometric phase $\theta_\text{G}$ jumps by $\pm\pi$. To understand the meaning of the jump, we recall that the Uhlmann parallel-transport condition is always satisfied in a Uhlmann quench, which ensures that $W(t+\dif t)$ remains as parallel as possible to $W(t)$ after the quench. However, the parallelism lacks transitivity \cite{OurPRB20b}, thereby $W(t)$ may lose parallelism with the initial amplitude $W(0)$. After crossing each $t^*_n$, the ``orientation'' of $W(t)$ changes from being parallel to being anti-parallel to $W(0)$ or vice versa, and the jump of the geometric phase reflects the change.
Thus, the occurrence of a GDQPT is accompanied by a change of the corresponding geometric phase associated with the evolution.
Additionally, the star symbols in the top panel of Figure~\ref{Fig1} indicate the times at which the geometric phase becomes the Uhlmann phase at $t=n\tau$ since the cyclic condition is satisfied. In this scenario, the time interval between adjacent GDQPTs is larger than $\tau$. The value of $\theta_U$ remains unchanged between two adjacent GDQPTs, as shown in the same panel.

In the limit when $T$ approaches $0$, $\beta\rightarrow +\infty$ and $\chi\rightarrow -1$, the Loschmidt amplitude reduces to a simple function: $\lim_{T\rightarrow 0}\mathcal{G}(T,t)=\cos\left(\frac{\omega_0t}{2}\right)$ according to Eq.~(\ref{tmp5c}). Consequently, both $\theta_\text{G}(t)$ and $r(t)$ act like periodic functions at $T=0.01\omega_0$. In this regime, $t^*_n\approx \frac{1}{\omega_0}(2n+1)\pi$ with $n$ being a non-negative integer. We illustrate the behaviors of $\theta_\text{G}(t)$ and $r(t)$ by the red solid lines respectively in the middle and bottom panels of Fig.~\ref{Fig1}. The black square symbols label where $\theta_\text{G}=\theta_U$ when the cyclic condition holds. Moreover, the period of $\theta_U$ is $\tau=\frac{2\pi}{\omega_0}\approx t^*_{n+1}-t^*_n$, implying that there is one square symbol between two adjacent GDQPTs in the middle panel.

The Uhlmann phase $\theta_U$ characterizes the topological nature of the evolution curve $W(t)$ via the Uhlmann holonomy in a cyclic process. At $T=0.01\omega_0$, the topological nature depends on whether the system undergoes odd (topological) or even (non-topological) times of the GDQPTs because of the twists of the holonomy. As illustrated in Fig.~\ref{Fig1}, $\theta_U=0$ ($\theta_U=\pi$) when there are odd (even) numbers of divergent points in $r(t)$ prior to $t$. In general, if $\theta_U(n\tau)=0$, then $W(n\tau)=W(0)$, indicating that the horizontal lift $\tilde{\gamma}$ of the evolution loop $\rho(t)$ is closed and the Uhlmann holonomy is trivial. If $\theta_U(n\tau)=\pi$, then $W(n\tau)\neq W(0)$, indicating that $\tilde{\gamma}$ is open and the Uhlmann holonomy is topologically nontrivial. The Uhlmann phase thus reflects the change of the Uhlmann holonomy.

\section{Experimental implications}\label{SecIV}
Experimental measurements of the Uhlmann phase remain a challenge, primarily due to the implementation of the Uhlmann parallel-transport condition which requires control of the ancilla representing the environment effect. A simulation of the Uhlmann phase of a simple two-level system was conducted on IBM's quantum computer \cite{npj18} with engineered Hamiltonians on both system and ancilla. Later, an experimental protocol based on the spin-$j$ systems was introduced in Ref.~\cite{OurPRA21}, where the Uhlmann parallel-transport condition is again achieved through unitary transformations on both system and ancilla.

Following these ideas, we may simulate the Uhlmann quench on quantum computers.
As an example, we take the system described in Sec.~\ref{SecIII} and choose the initial Hamiltonian as $H_0=\frac{\omega_0}{2}\sigma_z$ and the quench Hamiltonian as $H=\frac{\omega_0}{2}\left(\cos\phi\sigma_x+\sin\phi\sigma_y\right)$. In this case, a straightforward calculation shows $\tilde{H}=H$ according to the expression below Eq.~\eqref{dAU4}. After an Uhlmann quench, $W(t)$ evolves as
 \begin{align}
 W(t)=\me^{-\mi Ht}\sqrt{\rho(0)}\me^{\mi(1+\chi) Ht}
\end{align}
according to Eqs.~(\ref{Wtf}) and (\ref{gt}). The purification can be simulated on quantum computers by using the purified state
 \begin{align}\label{Wtb}
|W(t)\rangle=\sum_n\sqrt{\lambda_n}U_s(t)|n\rangle_s\otimes U^T_a(t)|n\rangle_a,
\end{align}
where $U_s(t)=\me^{-\mi Ht}$ and $U_a(t)=\me^{\mi(1+\chi) Ht}$ are unitary
operators acting on the system and ancilla states, respectively. $U_s(t)$ induces dynamic evolution of the system while $U_a(t)$ ensures the Uhlmann parallel-transport condition holds. $U_a(t)$ may be realized by determining the parameter $\chi$ from temperature via the expression below Eq.~\eqref{dAU4} and then constructing the effective evolution operator for the ancilla.
Once these two evolution operators are constructed, the Uhlmann quench may be simulated on the IBM Quantum Platform similar to that described in Ref.~\cite{npj18}.

Recently, significant progress has been made in experimental realizations of DQPTs on many different platforms, including the interacting transverse-field Ising
model \cite{DQPTB41}, topological nanomechanical systems \cite{PhysRevB.100.024310}, photonic platforms~\cite{WangPRL19}, superconducting qubits~\cite{GuoApplied19}, NV centers in diamonds~\cite{ChenAPL20}, and nuclear magnetic resonance quantum simulators~\cite{PhysRevLett.124.250601}. However, to the best of our knowledge, a direct measurement of the argument of the Loschmidt amplitude to infer the total phase has yet to be implemented. On the other hand, several techniques may help extract the features of the Uhlmann quench on quantum computers. For instance, the jumps of the Uhlmann phase may be detected through state tomography~\cite{npj18}, and the vanishing overlap of the purified states may be measured by quantum amplitude estimation~\cite{Brassard02}. Therefore, observation of the Uhlmann quench may be achieved on quantum computers by suitable transformations on the system and ancilla to generate the geometric phase in the dynamics after the quench.

\section{Conclusion}\label{SecV}
After examining and comparing the conditions rendering the Uhlmann parallel-transport and dynamics following a quench, we have developed a framework to integrate the geometric constraint into the quench dynamics within a single physical process called the Uhlmann quench. The incompatibility between the Uhlmann process and Hamiltonian dynamics is circumvented by introducing a more general evolution of the purification of the density matrix. The Uhlmann quench enables the generation of a geometric phase in the quench dynamics, which corresponds to the Uhlmann phase for systems with cyclic behavior of the density matrix. Moreover, GDQPTs with singular behavior can arise in the Uhlmann quench, as illustrated by a spin-$1/2$ system. With the rapid developing experimental progress on quantum quenches and DQPTs, our results pave a way for future experimental realization and observation of geometric behavior in quantum dynamics of mixed states.

\section{Acknowledgments}
We thank Dr. Kai-Yuan Cao for thoughtful discussions.
H.G. was supported by the National Natural Science
Foundation of China (Grant No. 12074064) and the Innovation Program for Quantum
Science and Technology (Grant No. 2021ZD0301904). X. Y. H. was supported by the Jiangsu Funding Program for Excellent Postdoctoral Talent (Grant No. 2023ZB611). C.C.C. was supported
by the National Science Foundation under Grant No.
PHY-2310656.

\appendix

\section{Fiber-bundle description of the Uhlmann phase}\label{appa00}
Let $\mathcal{P}$ be the space spanned by full-rank density matrices, which is also a manifold equipped with a Riemannian metric \cite{REP95}.
The associated purification form the surface of a unit sphere, $\mathcal{S}$, since $\langle W|W\rangle=\text{Tr}(W^\dag W)=\text{Tr}\rho=1$. After introducing the projection $\pi:\mathcal{S}\rightarrow \mathcal{P}$ such that $\pi(W)=WW^\dag=\rho$, a fiber bundle can be constructed in which $\mathcal{S}$ is the total space, and $\mathcal{P}$ is the base manifold. At each point $\rho\in \mathcal{P}$, $W=\sqrt{\rho}U$ spans the fiber $F_\rho$.
Clearly, $F_\rho\sim $U$(N)$, which is also isomorphic to the structural group, making the bundle a U$(N)$ principal bundle.

The parameter $\mathbf{R}$ can serve as the local coordinates of $\rho$. A continuous function $\sigma(\mathbf{R})\equiv\sigma(\rho(\mathbf{R}))=\sqrt{\rho(\mathbf{R})}U(\mathbf{R}) $ defines a section on the bundle. Since a global section $\sigma(\rho)=\sqrt{\rho}$ independent of the local coordinates always exists, this bundle is trivial~\cite{TDMPRB15}. On the base manifold $\mathcal{P}$, there exists a U$(N)$-gauge connection, known as the Uhlmann connection $A_U$, given by Eq.~(\ref{AU}). It is also the pull-back of the Ehresmann connection $\omega$ on the total space $\mathcal{S}$ via $\sigma$ \cite{ourPRB20}: $A_U=\sigma^*\omega$. $\omega$ separates the tangent bundle $T\mathcal{S}$ into its vertical and horizontal subspaces. We denote $\tilde{X}$ as the tangent vector of the curve $\tilde{\gamma}(t)$ given by Eq.~(\ref{rt}). Thus, the Uhlmann parallel-transport condition (\ref{pc}) can be expressed as~\cite{ourPRB20} $\omega(\tilde{X})=0$. In this case, $W(t)$ is called the horizontal lift of $\rho(t)=\pi(W(t))$. Since the solution of the first order differential equation (\ref{ptU}) is unique if the initial condition is given, every continuous curve on $\mathcal{P}$ has only one horizontal lift.

\section{Incompatibility between Uhlmann process and Hamiltonian dynamics}\label{appa0}
Here, we briefly review how the Uhlmann parallel-transport condition (\ref{pc}) is incompatible with the Hamiltonian dynamics governed by Eq.~(\ref{Wt2}).
By substituting the latter into the former, we obtain
\begin{align}\label{Wtf3}
HWW^\dag=-WW^\dag H,\quad \text{or} \quad \{H,\rho\}=0.
\end{align}
Since $\rho$ has full rank, its inverse exists. The above equation thereby leads to
\begin{align}
\sqrt{\rho^{-1}}H\sqrt{\rho}=-\sqrt{\rho}H\sqrt{\rho^{-1}}=-(\sqrt{\rho^{-1}}H\sqrt{\rho})^\dag,
\end{align}
where we have applied the fact that both $H$ and $\rho$ are Hermitian. We note that all eigenvalues of $\sqrt{\rho^{-1}}H\sqrt{\rho}$ are equal to those of $H$ since they differ only by a similarity transformation. Moreover, $H$ only have real eigenvalues, and Eq.~(\ref{Wtf3}) implies that all of its eigenvalues are zero, which is generally impossible. This proof applies to any dynamic process, regardless of whether it involves a quantum quench or not.

Ref.~\cite{Uhlmann91} considered a similar process in which $W$ evolves according to
 \begin{align}\label{Wt4}
\mi\dot{W}=H W-W\check{H},
\end{align}
where $\check{H}$ is referred to as the ``dual'' Hamiltonian. Obviously, if $\mi\mathcal{U}^\dag \dot{\mathcal{U}}=-\check{H}$, the generic evolution equation (\ref{Wt3}) recovers the ``dual'' dynamic equation (\ref{Wt4}).
However, Ref.~\cite{Uhlmann91} did not provide an explicit construction of $\check{H}$, a task accomplished in this work.

\section{Proof of the identity (\ref{tmp2d})}\label{appa}
The time-evolved density matrix has the form $\rho(t)=\sum_n\lambda_n(t)|n(t)\rangle\langle n(t)|$. Since $\me^{-\mi H t}$ is a unitary operator, $\lambda_n(t)=\lambda_n$, and $|n(t)\rangle=\me^{-\mi H t}|n\rangle$, $\rho(t)=\sum_n\lambda_n\me^{-\mi H t}|n\rangle\langle n|\me^{\mi H t}$. Moreover, $\sqrt{\rho(t)}=\sum_n\sqrt{\lambda_n}\me^{-\mi H t}|n\rangle\langle n|\me^{\mi H t}$, implying
 \begin{align}
\mi\dot{\sqrt{\rho(t)}}=[H,\sqrt{\rho(t)}].
\end{align}
By using Eq.~(\ref{AU}), Eq.~(\ref{tmp2d}) is equivalent to
\begin{widetext}
 \begin{align}\label{dAU}
\mi A_U(X)&=\sum_{nm}\me^{-\mi H t}|n\rangle\frac{\langle n|\me^{\mi H t}[\sqrt{\rho(t)},[H,\sqrt{\rho(t)}]]\me^{-\mi H t}|m\rangle}{\lambda_n+\lambda_m}\langle m|\me^{\mi H t}\notag\\
&=\sum_{nm}\me^{-\mi H t}|n\rangle\frac{\langle n|\me^{\mi H t}\left[2\sqrt{\rho(t)}H\sqrt{\rho(t)}-H\rho(t)-\rho(t)H\right]\me^{-\mi H t}|m\rangle}{\lambda_n+\lambda_m}\langle m|\me^{\mi H t}\notag\\
&=\sum_{nm}\me^{-\mi H t}|n\rangle\frac{\langle n|\left[2\sqrt{\rho(0)}H\sqrt{\rho(0)}-H\rho(0)-\rho(0)H\right]|m\rangle}{\lambda_n+\lambda_m}\langle m|\me^{\mi H t}\notag\\
&=\sum_{nm}\me^{-\mi H t}|n\rangle\frac{\langle n|\left[2\sqrt{\lambda_n\lambda_m}H-H(\lambda_n+\lambda_m)\right]|m\rangle}{\lambda_n+\lambda_m}\langle m|\me^{\mi H t}\notag\\
&=2\sum_{nm}\me^{-\mi H t}|n\rangle\frac{\langle n|\sqrt{\lambda_n\lambda_m}H|m\rangle}{\lambda_n+\lambda_m}\langle m|\me^{\mi H t}-H.
\end{align}
\end{widetext}
Substituting the expression into Eq.~(\ref{tmp2d}), we finally have
\begin{align}\label{tmp3}
\sqrt{\rho(0)} H \sqrt{\rho(0)}&=\{\rho(0),\sum_{nm}|n\rangle\frac{\langle n|\sqrt{\lambda_n\lambda_m}H|m\rangle}{\lambda_n+\lambda_m}\langle m| \}\notag\\
&=\sum_{nm}|n\rangle\langle n|\sqrt{\lambda_n\lambda_m}H|m\rangle\langle m|,
\end{align}
which is indeed an identity.

\bibliographystyle{apsrev}
\bibliography{Review1}

\begin{thebibliography}{64}
\expandafter\ifx\csname natexlab\endcsname\relax\def\natexlab#1{#1}\fi
\expandafter\ifx\csname bibnamefont\endcsname\relax
  \def\bibnamefont#1{#1}\fi
\expandafter\ifx\csname bibfnamefont\endcsname\relax
  \def\bibfnamefont#1{#1}\fi
\expandafter\ifx\csname citenamefont\endcsname\relax
  \def\citenamefont#1{#1}\fi
\expandafter\ifx\csname url\endcsname\relax
  \def\url#1{\texttt{#1}}\fi
\expandafter\ifx\csname urlprefix\endcsname\relax\def\urlprefix{URL }\fi
\providecommand{\bibinfo}[2]{#2}
\providecommand{\eprint}[2][]{\url{#2}}

\bibitem[{\citenamefont{Berry}(1984)}]{Berry84}
\bibinfo{author}{\bibfnamefont{M.~V.} \bibnamefont{Berry}},
  \bibinfo{journal}{Proc. R. Soc. A} \textbf{\bibinfo{volume}{392}},
  \bibinfo{pages}{45} (\bibinfo{year}{1984}).

\bibitem[{\citenamefont{Simon}(1983)}]{PhysRevLett.51.2167}
\bibinfo{author}{\bibfnamefont{B.}~\bibnamefont{Simon}},
  \bibinfo{journal}{Phys. Rev. Lett.} \textbf{\bibinfo{volume}{51}},
  \bibinfo{pages}{2167} (\bibinfo{year}{1983}),
  \urlprefix\url{https://link.aps.org/doi/10.1103/PhysRevLett.51.2167}.

\bibitem[{\citenamefont{Page}(1987)}]{PhysRevA.36.3479}
\bibinfo{author}{\bibfnamefont{D.~N.} \bibnamefont{Page}},
  \bibinfo{journal}{Phys. Rev. A} \textbf{\bibinfo{volume}{36}},
  \bibinfo{pages}{3479} (\bibinfo{year}{1987}),
  \urlprefix\url{https://link.aps.org/doi/10.1103/PhysRevA.36.3479}.

\bibitem[{\citenamefont{Cheng}(2010)}]{QGT10}
\bibinfo{author}{\bibfnamefont{R.}~\bibnamefont{Cheng}},
  \emph{\bibinfo{title}{Quantum geometric tensor (fubini-study metric) in
  simple quantum system: A pedagogical introduction}} (\bibinfo{year}{2010}),
  \bibinfo{note}{arXiv:1012.1337}.

\bibitem[{\citenamefont{Bohm et~al.}(2003)\citenamefont{Bohm, Mostafazadeh,
  Koizumi, Niu, and Zwanziger}}]{Bohm03}
\bibinfo{author}{\bibfnamefont{A.}~\bibnamefont{Bohm}},
  \bibinfo{author}{\bibfnamefont{A.}~\bibnamefont{Mostafazadeh}},
  \bibinfo{author}{\bibfnamefont{H.}~\bibnamefont{Koizumi}},
  \bibinfo{author}{\bibfnamefont{Q.}~\bibnamefont{Niu}}, \bibnamefont{and}
  \bibinfo{author}{\bibfnamefont{J.}~\bibnamefont{Zwanziger}},
  \emph{\bibinfo{title}{The geometric phase in quantum systems}}
  (\bibinfo{publisher}{Springer}, \bibinfo{address}{Berlin, Germany},
  \bibinfo{year}{2003}).

\bibitem[{\citenamefont{Vanderbilt}(2018)}]{Vanderbilt_book}
\bibinfo{author}{\bibfnamefont{D.}~\bibnamefont{Vanderbilt}},
  \emph{\bibinfo{title}{Berry Phases in Electronic Structure Theory: Electric
  Polarization, Orbital Magnetization and Topological Insulators}}
  (\bibinfo{publisher}{Cambridge University Press},
  \bibinfo{address}{Cambridge, UK}, \bibinfo{year}{2018}).

\bibitem[{\citenamefont{Cohen et~al.}(2019)\citenamefont{Cohen, Larocque,
  Bouchard, Nejadsattari, Gefen, and Karimi}}]{Cohen19}
\bibinfo{author}{\bibfnamefont{E.}~\bibnamefont{Cohen}},
  \bibinfo{author}{\bibfnamefont{H.}~\bibnamefont{Larocque}},
  \bibinfo{author}{\bibfnamefont{F.}~\bibnamefont{Bouchard}},
  \bibinfo{author}{\bibfnamefont{F.}~\bibnamefont{Nejadsattari}},
  \bibinfo{author}{\bibfnamefont{Y.}~\bibnamefont{Gefen}}, \bibnamefont{and}
  \bibinfo{author}{\bibfnamefont{E.}~\bibnamefont{Karimi}},
  \bibinfo{journal}{Nat. Rev. Phys.} \textbf{\bibinfo{volume}{1}},
  \bibinfo{pages}{437} (\bibinfo{year}{2019}).

\bibitem[{\citenamefont{Kitaev}(2001)}]{Kitaev_2001}
\bibinfo{author}{\bibfnamefont{A.~Y.} \bibnamefont{Kitaev}},
  \bibinfo{journal}{Physics-Uspekhi} \textbf{\bibinfo{volume}{44}},
  \bibinfo{pages}{131} (\bibinfo{year}{2001}),
  \urlprefix\url{https://dx.doi.org/10.1070/1063-7869/44/10S/S29}.

\bibitem[{\citenamefont{Nayak et~al.}(2008)\citenamefont{Nayak, Simon, Stern,
  Freedman, and Das~Sarma}}]{RevModPhys.80.1083}
\bibinfo{author}{\bibfnamefont{C.}~\bibnamefont{Nayak}},
  \bibinfo{author}{\bibfnamefont{S.~H.} \bibnamefont{Simon}},
  \bibinfo{author}{\bibfnamefont{A.}~\bibnamefont{Stern}},
  \bibinfo{author}{\bibfnamefont{M.}~\bibnamefont{Freedman}}, \bibnamefont{and}
  \bibinfo{author}{\bibfnamefont{S.}~\bibnamefont{Das~Sarma}},
  \bibinfo{journal}{Rev. Mod. Phys.} \textbf{\bibinfo{volume}{80}},
  \bibinfo{pages}{1083} (\bibinfo{year}{2008}),
  \urlprefix\url{https://link.aps.org/doi/10.1103/RevModPhys.80.1083}.

\bibitem[{\citenamefont{Alicea et~al.}(2011)\citenamefont{Alicea, Oreg, Refael,
  von Oppen, and Fisher}}]{AliceaNP11}
\bibinfo{author}{\bibfnamefont{J.}~\bibnamefont{Alicea}},
  \bibinfo{author}{\bibfnamefont{Y.}~\bibnamefont{Oreg}},
  \bibinfo{author}{\bibfnamefont{G.}~\bibnamefont{Refael}},
  \bibinfo{author}{\bibfnamefont{F.}~\bibnamefont{von Oppen}},
  \bibnamefont{and} \bibinfo{author}{\bibfnamefont{M.~P.~A.}
  \bibnamefont{Fisher}}, \bibinfo{journal}{Nat. Phys.}
  \textbf{\bibinfo{volume}{7}}, \bibinfo{pages}{412} (\bibinfo{year}{2011}).

\bibitem[{\citenamefont{Lahtinen and Pachos}(2017)}]{SciPostPhys.3.3.021}
\bibinfo{author}{\bibfnamefont{V.}~\bibnamefont{Lahtinen}} \bibnamefont{and}
  \bibinfo{author}{\bibfnamefont{J.~K.} \bibnamefont{Pachos}},
  \bibinfo{journal}{SciPost Phys.} \textbf{\bibinfo{volume}{3}},
  \bibinfo{pages}{021} (\bibinfo{year}{2017}),
  \urlprefix\url{https://scipost.org/10.21468/SciPostPhys.3.3.021}.

\bibitem[{\citenamefont{Hasan and Kane}(2010)}]{KaneRMP}
\bibinfo{author}{\bibfnamefont{M.~Z.} \bibnamefont{Hasan}} \bibnamefont{and}
  \bibinfo{author}{\bibfnamefont{C.~L.} \bibnamefont{Kane}},
  \bibinfo{journal}{Rev. Mod. Phys.} \textbf{\bibinfo{volume}{82}},
  \bibinfo{pages}{3045} (\bibinfo{year}{2010}).

\bibitem[{\citenamefont{Qi and Zhang}(2011)}]{ZhangSCRMP}
\bibinfo{author}{\bibfnamefont{X.-L.} \bibnamefont{Qi}} \bibnamefont{and}
  \bibinfo{author}{\bibfnamefont{S.-C.} \bibnamefont{Zhang}},
  \bibinfo{journal}{Rev. Mod. Phys.} \textbf{\bibinfo{volume}{83}},
  \bibinfo{pages}{1057} (\bibinfo{year}{2011}).

\bibitem[{\citenamefont{Bernevig and Hughes}(2013)}]{Bernevigbook}
\bibinfo{author}{\bibfnamefont{B.~A.} \bibnamefont{Bernevig}} \bibnamefont{and}
  \bibinfo{author}{\bibfnamefont{T.~L.} \bibnamefont{Hughes}},
  \emph{\bibinfo{title}{Topological Insulators and Topological
  Superconductors}} (\bibinfo{publisher}{Princeton, NJ}, \bibinfo{year}{2013}).

\bibitem[{\citenamefont{Chiu et~al.}(2016)\citenamefont{Chiu, Teo, Schnyder,
  and Ryu}}]{ChiuRMP}
\bibinfo{author}{\bibfnamefont{C.~K.} \bibnamefont{Chiu}},
  \bibinfo{author}{\bibfnamefont{J.~C.~Y.} \bibnamefont{Teo}},
  \bibinfo{author}{\bibfnamefont{A.~P.} \bibnamefont{Schnyder}},
  \bibnamefont{and} \bibinfo{author}{\bibfnamefont{S.}~\bibnamefont{Ryu}},
  \bibinfo{journal}{Rev. Mod. Phys.} \textbf{\bibinfo{volume}{88}},
  \bibinfo{pages}{035005} (\bibinfo{year}{2016}).

\bibitem[{\citenamefont{Moore}(2010)}]{MooreN}
\bibinfo{author}{\bibfnamefont{J.~E.} \bibnamefont{Moore}},
  \bibinfo{journal}{Nature} \textbf{\bibinfo{volume}{464}},
  \bibinfo{pages}{194} (\bibinfo{year}{2010}).

\bibitem[{\citenamefont{Kane and Mele}(2005{\natexlab{a}})}]{KaneMele}
\bibinfo{author}{\bibfnamefont{C.~L.} \bibnamefont{Kane}} \bibnamefont{and}
  \bibinfo{author}{\bibfnamefont{E.~J.} \bibnamefont{Mele}},
  \bibinfo{journal}{Phys. Rev. Lett.} \textbf{\bibinfo{volume}{95}},
  \bibinfo{pages}{226801} (\bibinfo{year}{2005}{\natexlab{a}}).

\bibitem[{\citenamefont{Kane and Mele}(2005{\natexlab{b}})}]{KaneMele2}
\bibinfo{author}{\bibfnamefont{C.~L.} \bibnamefont{Kane}} \bibnamefont{and}
  \bibinfo{author}{\bibfnamefont{E.~J.} \bibnamefont{Mele}},
  \bibinfo{journal}{Phys. Rev. Lett.} \textbf{\bibinfo{volume}{95}},
  \bibinfo{pages}{146802} (\bibinfo{year}{2005}{\natexlab{b}}).

\bibitem[{\citenamefont{Bernevig and Zhang}(2006)}]{BernevigPRL}
\bibinfo{author}{\bibfnamefont{B.~A.} \bibnamefont{Bernevig}} \bibnamefont{and}
  \bibinfo{author}{\bibfnamefont{S.-C.} \bibnamefont{Zhang}},
  \bibinfo{journal}{Phys. Rev. Lett.} \textbf{\bibinfo{volume}{96}},
  \bibinfo{pages}{106802} (\bibinfo{year}{2006}).

\bibitem[{\citenamefont{Moore and Balents}(2007)}]{MoorePRB}
\bibinfo{author}{\bibfnamefont{J.~E.} \bibnamefont{Moore}} \bibnamefont{and}
  \bibinfo{author}{\bibfnamefont{L.}~\bibnamefont{Balents}},
  \bibinfo{journal}{Phys. Rev. B} \textbf{\bibinfo{volume}{75}},
  \bibinfo{pages}{121306(R)} (\bibinfo{year}{2007}).

\bibitem[{\citenamefont{Fu et~al.}(2007)\citenamefont{Fu, Kane, and
  Mele}}]{FuLPRL}
\bibinfo{author}{\bibfnamefont{L.}~\bibnamefont{Fu}},
  \bibinfo{author}{\bibfnamefont{C.~L.} \bibnamefont{Kane}}, \bibnamefont{and}
  \bibinfo{author}{\bibfnamefont{E.~J.} \bibnamefont{Mele}},
  \bibinfo{journal}{Phys. Rev. Lett.} \textbf{\bibinfo{volume}{98}},
  \bibinfo{pages}{106803} (\bibinfo{year}{2007}).

\bibitem[{\citenamefont{Thouless et~al.}(1982)\citenamefont{Thouless, Kohmoto,
  Nightingale, and den Nijs}}]{TKNN}
\bibinfo{author}{\bibfnamefont{D.~J.} \bibnamefont{Thouless}},
  \bibinfo{author}{\bibfnamefont{M.}~\bibnamefont{Kohmoto}},
  \bibinfo{author}{\bibfnamefont{M.~P.} \bibnamefont{Nightingale}},
  \bibnamefont{and} \bibinfo{author}{\bibfnamefont{M.}~\bibnamefont{den Nijs}},
  \bibinfo{journal}{Phys. Rev. Lett.} \textbf{\bibinfo{volume}{49}},
  \bibinfo{pages}{405} (\bibinfo{year}{1982}).

\bibitem[{\citenamefont{Haldane}(1988)}]{Haldane}
\bibinfo{author}{\bibfnamefont{F.~D.~M.} \bibnamefont{Haldane}},
  \bibinfo{journal}{Phys. Rev. Lett.} \textbf{\bibinfo{volume}{61}},
  \bibinfo{pages}{2015} (\bibinfo{year}{1988}).

\bibitem[{\citenamefont{Xiao et~al.}(2010)\citenamefont{Xiao, Chang, and
  Niu}}]{RevModPhys.82.1959}
\bibinfo{author}{\bibfnamefont{D.}~\bibnamefont{Xiao}},
  \bibinfo{author}{\bibfnamefont{M.-C.} \bibnamefont{Chang}}, \bibnamefont{and}
  \bibinfo{author}{\bibfnamefont{Q.}~\bibnamefont{Niu}}, \bibinfo{journal}{Rev.
  Mod. Phys.} \textbf{\bibinfo{volume}{82}}, \bibinfo{pages}{1959}
  (\bibinfo{year}{2010}),
  \urlprefix\url{https://link.aps.org/doi/10.1103/RevModPhys.82.1959}.

\bibitem[{\citenamefont{Uhlmann}(1986)}]{Uhlmann86}
\bibinfo{author}{\bibfnamefont{A.}~\bibnamefont{Uhlmann}},
  \bibinfo{journal}{Rep. Math. Phys.} \textbf{\bibinfo{volume}{24}},
  \bibinfo{pages}{229} (\bibinfo{year}{1986}).

\bibitem[{\citenamefont{Uhlmann}(1989)}]{Uhlmann89}
\bibinfo{author}{\bibfnamefont{A.}~\bibnamefont{Uhlmann}},
  \bibinfo{journal}{Ann. Phys. (Berlin)} \textbf{\bibinfo{volume}{501}},
  \bibinfo{pages}{63} (\bibinfo{year}{1989}).

\bibitem[{\citenamefont{Uhlmann}(1991)}]{Uhlmann91}
\bibinfo{author}{\bibfnamefont{A.}~\bibnamefont{Uhlmann}},
  \bibinfo{journal}{Lett. Math. Phys.} \textbf{\bibinfo{volume}{21}},
  \bibinfo{pages}{229} (\bibinfo{year}{1991}).

\bibitem[{\citenamefont{Sj\"oqvist et~al.}(2000)\citenamefont{Sj\"oqvist, Pati,
  Ekert, Anandan, Ericsson, Oi, and Vedral}}]{PhysRevLett.85.2845}
\bibinfo{author}{\bibfnamefont{E.}~\bibnamefont{Sj\"oqvist}},
  \bibinfo{author}{\bibfnamefont{A.~K.} \bibnamefont{Pati}},
  \bibinfo{author}{\bibfnamefont{A.}~\bibnamefont{Ekert}},
  \bibinfo{author}{\bibfnamefont{J.~S.} \bibnamefont{Anandan}},
  \bibinfo{author}{\bibfnamefont{M.}~\bibnamefont{Ericsson}},
  \bibinfo{author}{\bibfnamefont{D.~K.~L.} \bibnamefont{Oi}}, \bibnamefont{and}
  \bibinfo{author}{\bibfnamefont{V.}~\bibnamefont{Vedral}},
  \bibinfo{journal}{Phys. Rev. Lett.} \textbf{\bibinfo{volume}{85}},
  \bibinfo{pages}{2845} (\bibinfo{year}{2000}),
  \urlprefix\url{https://link.aps.org/doi/10.1103/PhysRevLett.85.2845}.

\bibitem[{\citenamefont{Budich and
  Diehl}(2015{\natexlab{a}})}]{PhysRevB.91.165140}
\bibinfo{author}{\bibfnamefont{J.~C.} \bibnamefont{Budich}} \bibnamefont{and}
  \bibinfo{author}{\bibfnamefont{S.}~\bibnamefont{Diehl}},
  \bibinfo{journal}{Phys. Rev. B} \textbf{\bibinfo{volume}{91}},
  \bibinfo{pages}{165140} (\bibinfo{year}{2015}{\natexlab{a}}),
  \urlprefix\url{https://link.aps.org/doi/10.1103/PhysRevB.91.165140}.

\bibitem[{\citenamefont{Guo et~al.}(2020)\citenamefont{Guo, Hou, He, and
  Chien}}]{ourPRB20}
\bibinfo{author}{\bibfnamefont{H.}~\bibnamefont{Guo}},
  \bibinfo{author}{\bibfnamefont{X.-Y.} \bibnamefont{Hou}},
  \bibinfo{author}{\bibfnamefont{Y.}~\bibnamefont{He}}, \bibnamefont{and}
  \bibinfo{author}{\bibfnamefont{C.~C.} \bibnamefont{Chien}},
  \bibinfo{journal}{Phys. Rev. B} \textbf{\bibinfo{volume}{101}},
  \bibinfo{pages}{104310} (\bibinfo{year}{2020}).

\bibitem[{\citenamefont{Viyuela et~al.}(2014)\citenamefont{Viyuela, Rivas, and
  Martin-Delgado}}]{PhysRevLett.112.130401}
\bibinfo{author}{\bibfnamefont{O.}~\bibnamefont{Viyuela}},
  \bibinfo{author}{\bibfnamefont{A.}~\bibnamefont{Rivas}}, \bibnamefont{and}
  \bibinfo{author}{\bibfnamefont{M.~A.} \bibnamefont{Martin-Delgado}},
  \bibinfo{journal}{Phys. Rev. Lett.} \textbf{\bibinfo{volume}{112}},
  \bibinfo{pages}{130401} (\bibinfo{year}{2014}),
  \urlprefix\url{https://link.aps.org/doi/10.1103/PhysRevLett.112.130401}.

\bibitem[{\citenamefont{Huang and Arovas}(2014)}]{PhysRevLett.113.076407}
\bibinfo{author}{\bibfnamefont{Z.}~\bibnamefont{Huang}} \bibnamefont{and}
  \bibinfo{author}{\bibfnamefont{D.~P.} \bibnamefont{Arovas}},
  \bibinfo{journal}{Phys. Rev. Lett.} \textbf{\bibinfo{volume}{113}},
  \bibinfo{pages}{076407} (\bibinfo{year}{2014}),
  \urlprefix\url{https://link.aps.org/doi/10.1103/PhysRevLett.113.076407}.

\bibitem[{\citenamefont{Carollo et~al.}(2018)\citenamefont{Carollo, Spagnolo,
  and Valenti}}]{Uhlmann17}
\bibinfo{author}{\bibfnamefont{A.}~\bibnamefont{Carollo}},
  \bibinfo{author}{\bibfnamefont{B.}~\bibnamefont{Spagnolo}}, \bibnamefont{and}
  \bibinfo{author}{\bibfnamefont{D.}~\bibnamefont{Valenti}},
  \bibinfo{journal}{Sci. Rep.} \textbf{\bibinfo{volume}{8}},
  \bibinfo{pages}{9852} (\bibinfo{year}{2018}).

\bibitem[{\citenamefont{Leonforte et~al.}(2019)\citenamefont{Leonforte,
  Valenti, Spagnolo, and Carollo}}]{Uhlmann18}
\bibinfo{author}{\bibfnamefont{L.}~\bibnamefont{Leonforte}},
  \bibinfo{author}{\bibfnamefont{D.}~\bibnamefont{Valenti}},
  \bibinfo{author}{\bibfnamefont{B.}~\bibnamefont{Spagnolo}}, \bibnamefont{and}
  \bibinfo{author}{\bibfnamefont{A.}~\bibnamefont{Carollo}},
  \bibinfo{journal}{Sci. Rep.} \textbf{\bibinfo{volume}{9}},
  \bibinfo{pages}{9106} (\bibinfo{year}{2019}).

\bibitem[{\citenamefont{Viyuela et~al.}(2018)\citenamefont{Viyuela, Rivas,
  Gasparinetti, Wallraff, Filipp, and Martin-Delgado}}]{npj18}
\bibinfo{author}{\bibfnamefont{O.}~\bibnamefont{Viyuela}},
  \bibinfo{author}{\bibfnamefont{A.}~\bibnamefont{Rivas}},
  \bibinfo{author}{\bibfnamefont{S.}~\bibnamefont{Gasparinetti}},
  \bibinfo{author}{\bibfnamefont{A.}~\bibnamefont{Wallraff}},
  \bibinfo{author}{\bibfnamefont{S.}~\bibnamefont{Filipp}}, \bibnamefont{and}
  \bibinfo{author}{\bibfnamefont{M.~A.} \bibnamefont{Martin-Delgado}},
  \bibinfo{journal}{npj Quant. Inf.} \textbf{\bibinfo{volume}{4}},
  \bibinfo{pages}{10} (\bibinfo{year}{2018}).

\bibitem[{\citenamefont{Mera et~al.}(2017)\citenamefont{Mera, Vlachou,
  Paunkovi\ifmmode~\acute{c}\else \'{c}\fi{}, and
  Vieira}}]{PhysRevLett.119.015702}
\bibinfo{author}{\bibfnamefont{B.}~\bibnamefont{Mera}},
  \bibinfo{author}{\bibfnamefont{C.}~\bibnamefont{Vlachou}},
  \bibinfo{author}{\bibfnamefont{N.}~\bibnamefont{Paunkovi\ifmmode~\acute{c}\else
  \'{c}\fi{}}}, \bibnamefont{and} \bibinfo{author}{\bibfnamefont{V.~R.}
  \bibnamefont{Vieira}}, \bibinfo{journal}{Phys. Rev. Lett.}
  \textbf{\bibinfo{volume}{119}}, \bibinfo{pages}{015702}
  (\bibinfo{year}{2017}),
  \urlprefix\url{https://link.aps.org/doi/10.1103/PhysRevLett.119.015702}.

\bibitem[{\citenamefont{He et~al.}(2018)\citenamefont{He, Guo, and
  Chien}}]{PhysRevB.97.235141}
\bibinfo{author}{\bibfnamefont{Y.}~\bibnamefont{He}},
  \bibinfo{author}{\bibfnamefont{H.}~\bibnamefont{Guo}}, \bibnamefont{and}
  \bibinfo{author}{\bibfnamefont{C.-C.} \bibnamefont{Chien}},
  \bibinfo{journal}{Phys. Rev. B} \textbf{\bibinfo{volume}{97}},
  \bibinfo{pages}{235141} (\bibinfo{year}{2018}),
  \urlprefix\url{https://link.aps.org/doi/10.1103/PhysRevB.97.235141}.

\bibitem[{\citenamefont{Hauru and Vidal}(2018)}]{PhysRevA.98.042316}
\bibinfo{author}{\bibfnamefont{M.}~\bibnamefont{Hauru}} \bibnamefont{and}
  \bibinfo{author}{\bibfnamefont{G.}~\bibnamefont{Vidal}},
  \bibinfo{journal}{Phys. Rev. A} \textbf{\bibinfo{volume}{98}},
  \bibinfo{pages}{042316} (\bibinfo{year}{2018}),
  \urlprefix\url{https://link.aps.org/doi/10.1103/PhysRevA.98.042316}.

\bibitem[{\citenamefont{Hou et~al.}(2020)\citenamefont{Hou, Gao, Guo, He, Liu,
  and Chien}}]{OurPRB20b}
\bibinfo{author}{\bibfnamefont{X.-Y.} \bibnamefont{Hou}},
  \bibinfo{author}{\bibfnamefont{Q.-C.} \bibnamefont{Gao}},
  \bibinfo{author}{\bibfnamefont{H.}~\bibnamefont{Guo}},
  \bibinfo{author}{\bibfnamefont{Y.}~\bibnamefont{He}},
  \bibinfo{author}{\bibfnamefont{T.}~\bibnamefont{Liu}}, \bibnamefont{and}
  \bibinfo{author}{\bibfnamefont{C.~C.} \bibnamefont{Chien}},
  \bibinfo{journal}{Phys. Rev. B} \textbf{\bibinfo{volume}{102}},
  \bibinfo{pages}{104305} (\bibinfo{year}{2020}).

\bibitem[{\citenamefont{Morachis~Galindo
  et~al.}(2021)\citenamefont{Morachis~Galindo, Rojas, and
  Maytorena}}]{Galindo21}
\bibinfo{author}{\bibfnamefont{D.}~\bibnamefont{Morachis~Galindo}},
  \bibinfo{author}{\bibfnamefont{F.}~\bibnamefont{Rojas}}, \bibnamefont{and}
  \bibinfo{author}{\bibfnamefont{J.~A.} \bibnamefont{Maytorena}},
  \bibinfo{journal}{Phys. Rev. A} \textbf{\bibinfo{volume}{103}},
  \bibinfo{pages}{042221} (\bibinfo{year}{2021}).

\bibitem[{\citenamefont{Hou et~al.}(2023)\citenamefont{Hou, Zhou, Wang, Guo,
  and Chien}}]{OurQGT23}
\bibinfo{author}{\bibfnamefont{X.-Y.} \bibnamefont{Hou}},
  \bibinfo{author}{\bibfnamefont{Z.}~\bibnamefont{Zhou}},
  \bibinfo{author}{\bibfnamefont{X.}~\bibnamefont{Wang}},
  \bibinfo{author}{\bibfnamefont{H.}~\bibnamefont{Guo}}, \bibnamefont{and}
  \bibinfo{author}{\bibfnamefont{C.-C.} \bibnamefont{Chien}},
  \emph{\bibinfo{title}{Local geometry and quantum geometric tensor of mixed
  states}} (\bibinfo{year}{2023}), \bibinfo{note}{arXiv:2305.07597}.

\bibitem[{\citenamefont{Hou et~al.}(2021)\citenamefont{Hou, Guo, and
  Chien}}]{OurPRA21}
\bibinfo{author}{\bibfnamefont{X.-Y.} \bibnamefont{Hou}},
  \bibinfo{author}{\bibfnamefont{H.}~\bibnamefont{Guo}}, \bibnamefont{and}
  \bibinfo{author}{\bibfnamefont{C.~C.} \bibnamefont{Chien}},
  \bibinfo{journal}{Phys. Rev. A} \textbf{\bibinfo{volume}{104}},
  \bibinfo{pages}{023303} (\bibinfo{year}{2021}).

\bibitem[{\citenamefont{Heyl}(2014)}]{DQPT14}
\bibinfo{author}{\bibfnamefont{M.}~\bibnamefont{Heyl}}, \bibinfo{journal}{Phys.
  Rev. Lett.} \textbf{\bibinfo{volume}{113}}, \bibinfo{pages}{205701}
  (\bibinfo{year}{2014}).

\bibitem[{\citenamefont{Heyl}(2015)}]{DQPT15}
\bibinfo{author}{\bibfnamefont{M.}~\bibnamefont{Heyl}}, \bibinfo{journal}{Phys.
  Rev. Lett.} \textbf{\bibinfo{volume}{115}}, \bibinfo{pages}{140602}
  (\bibinfo{year}{2015}).

\bibitem[{\citenamefont{Sharma et~al.}(2016)\citenamefont{Sharma, Divakaran,
  Polkovnikov, and Dutta}}]{DQPTB2}
\bibinfo{author}{\bibfnamefont{S.}~\bibnamefont{Sharma}},
  \bibinfo{author}{\bibfnamefont{U.}~\bibnamefont{Divakaran}},
  \bibinfo{author}{\bibfnamefont{A.}~\bibnamefont{Polkovnikov}},
  \bibnamefont{and} \bibinfo{author}{\bibfnamefont{A.}~\bibnamefont{Dutta}},
  \bibinfo{journal}{Phys. Rev. B} \textbf{\bibinfo{volume}{93}},
  \bibinfo{pages}{144306} (\bibinfo{year}{2016}).

\bibitem[{\citenamefont{Budich and Heyl}(2016{\natexlab{a}})}]{DQPTPRB16}
\bibinfo{author}{\bibfnamefont{J.~C.} \bibnamefont{Budich}} \bibnamefont{and}
  \bibinfo{author}{\bibfnamefont{M.}~\bibnamefont{Heyl}},
  \bibinfo{journal}{Phys. Rev. B} \textbf{\bibinfo{volume}{93}},
  \bibinfo{pages}{085416} (\bibinfo{year}{2016}{\natexlab{a}}).

\bibitem[{\citenamefont{Heyl}(2018)}]{DQPTreview18}
\bibinfo{author}{\bibfnamefont{M.}~\bibnamefont{Heyl}}, \bibinfo{journal}{Rep.
  Prog. Phys.} \textbf{\bibinfo{volume}{81}}, \bibinfo{pages}{054001}
  (\bibinfo{year}{2018}).

\bibitem[{\citenamefont{Jurcevic et~al.}(2017)\citenamefont{Jurcevic, Shen,
  Hauke, Maier, Brydges, Hempel, Lanyon, Heyl, Blatt, and Roos}}]{DQPTB41}
\bibinfo{author}{\bibfnamefont{P.}~\bibnamefont{Jurcevic}},
  \bibinfo{author}{\bibfnamefont{H.}~\bibnamefont{Shen}},
  \bibinfo{author}{\bibfnamefont{P.}~\bibnamefont{Hauke}},
  \bibinfo{author}{\bibfnamefont{C.}~\bibnamefont{Maier}},
  \bibinfo{author}{\bibfnamefont{T.}~\bibnamefont{Brydges}},
  \bibinfo{author}{\bibfnamefont{C.}~\bibnamefont{Hempel}},
  \bibinfo{author}{\bibfnamefont{B.~P.} \bibnamefont{Lanyon}},
  \bibinfo{author}{\bibfnamefont{M.}~\bibnamefont{Heyl}},
  \bibinfo{author}{\bibfnamefont{R.}~\bibnamefont{Blatt}}, \bibnamefont{and}
  \bibinfo{author}{\bibfnamefont{C.~F.} \bibnamefont{Roos}},
  \bibinfo{journal}{Phys. Rev. Lett.} \textbf{\bibinfo{volume}{119}},
  \bibinfo{pages}{080501} (\bibinfo{year}{2017}).

\bibitem[{\citenamefont{Fl$\ddot{\textrm{a}}$schner
  et~al.}(2018)\citenamefont{Fl$\ddot{\textrm{a}}$schner, Vogel, Tarnowski,
  Rem, Luhmann, Heyl, Budich, Mathe, Sengstock, and Weitenberg}}]{DQPTB4}
\bibinfo{author}{\bibfnamefont{N.}~\bibnamefont{Fl$\ddot{\textrm{a}}$schner}},
  \bibinfo{author}{\bibfnamefont{D.}~\bibnamefont{Vogel}},
  \bibinfo{author}{\bibfnamefont{M.}~\bibnamefont{Tarnowski}},
  \bibinfo{author}{\bibfnamefont{B.~S.} \bibnamefont{Rem}},
  \bibinfo{author}{\bibfnamefont{D.~S.} \bibnamefont{Luhmann}},
  \bibinfo{author}{\bibfnamefont{M.}~\bibnamefont{Heyl}},
  \bibinfo{author}{\bibfnamefont{J.~C.} \bibnamefont{Budich}},
  \bibinfo{author}{\bibfnamefont{Y.~L.} \bibnamefont{Mathe}},
  \bibinfo{author}{\bibfnamefont{K.}~\bibnamefont{Sengstock}},
  \bibnamefont{and}
  \bibinfo{author}{\bibfnamefont{C.}~\bibnamefont{Weitenberg}},
  \bibinfo{journal}{Nat. Phys.} \textbf{\bibinfo{volume}{14}},
  \bibinfo{pages}{265} (\bibinfo{year}{2018}).

\bibitem[{\citenamefont{Zhang et~al.}(2017)\citenamefont{Zhang, Pagano, Hess,
  Kyprianidis, Becker, Kaplan, Gorshkov, Gong, and Monroe}}]{Zhang_2017}
\bibinfo{author}{\bibfnamefont{J.}~\bibnamefont{Zhang}},
  \bibinfo{author}{\bibfnamefont{G.}~\bibnamefont{Pagano}},
  \bibinfo{author}{\bibfnamefont{P.~W.} \bibnamefont{Hess}},
  \bibinfo{author}{\bibfnamefont{A.}~\bibnamefont{Kyprianidis}},
  \bibinfo{author}{\bibfnamefont{P.}~\bibnamefont{Becker}},
  \bibinfo{author}{\bibfnamefont{H.}~\bibnamefont{Kaplan}},
  \bibinfo{author}{\bibfnamefont{A.~V.} \bibnamefont{Gorshkov}},
  \bibinfo{author}{\bibfnamefont{Z.-X.} \bibnamefont{Gong}}, \bibnamefont{and}
  \bibinfo{author}{\bibfnamefont{C.}~\bibnamefont{Monroe}},
  \bibinfo{journal}{Nature} \textbf{\bibinfo{volume}{551}},
  \bibinfo{pages}{601–604} (\bibinfo{year}{2017}), ISSN
  \bibinfo{issn}{1476-4687},
  \urlprefix\url{http://dx.doi.org/10.1038/nature24654}.

\bibitem[{\citenamefont{Guo et~al.}(2019{\natexlab{a}})\citenamefont{Guo, Yang,
  Zeng, Peng, Li, Deng, Jin, Chen, Zheng, and Fan}}]{PhysRevApplied.11.044080}
\bibinfo{author}{\bibfnamefont{X.-Y.} \bibnamefont{Guo}},
  \bibinfo{author}{\bibfnamefont{C.}~\bibnamefont{Yang}},
  \bibinfo{author}{\bibfnamefont{Y.}~\bibnamefont{Zeng}},
  \bibinfo{author}{\bibfnamefont{Y.}~\bibnamefont{Peng}},
  \bibinfo{author}{\bibfnamefont{H.-K.} \bibnamefont{Li}},
  \bibinfo{author}{\bibfnamefont{H.}~\bibnamefont{Deng}},
  \bibinfo{author}{\bibfnamefont{Y.-R.} \bibnamefont{Jin}},
  \bibinfo{author}{\bibfnamefont{S.}~\bibnamefont{Chen}},
  \bibinfo{author}{\bibfnamefont{D.}~\bibnamefont{Zheng}}, \bibnamefont{and}
  \bibinfo{author}{\bibfnamefont{H.}~\bibnamefont{Fan}},
  \bibinfo{journal}{Phys. Rev. Appl.} \textbf{\bibinfo{volume}{11}},
  \bibinfo{pages}{044080} (\bibinfo{year}{2019}{\natexlab{a}}),
  \urlprefix\url{https://link.aps.org/doi/10.1103/PhysRevApplied.11.044080}.

\bibitem[{\citenamefont{Wang et~al.}(2019{\natexlab{a}})\citenamefont{Wang,
  Qiu, Xiao, Zhan, Bian, Yi, and Xue}}]{PhysRevLett.122.020501}
\bibinfo{author}{\bibfnamefont{K.}~\bibnamefont{Wang}},
  \bibinfo{author}{\bibfnamefont{X.}~\bibnamefont{Qiu}},
  \bibinfo{author}{\bibfnamefont{L.}~\bibnamefont{Xiao}},
  \bibinfo{author}{\bibfnamefont{X.}~\bibnamefont{Zhan}},
  \bibinfo{author}{\bibfnamefont{Z.}~\bibnamefont{Bian}},
  \bibinfo{author}{\bibfnamefont{W.}~\bibnamefont{Yi}}, \bibnamefont{and}
  \bibinfo{author}{\bibfnamefont{P.}~\bibnamefont{Xue}},
  \bibinfo{journal}{Phys. Rev. Lett.} \textbf{\bibinfo{volume}{122}},
  \bibinfo{pages}{020501} (\bibinfo{year}{2019}{\natexlab{a}}),
  \urlprefix\url{https://link.aps.org/doi/10.1103/PhysRevLett.122.020501}.

\bibitem[{\citenamefont{Tian et~al.}(2019)\citenamefont{Tian, Ke, Zhang, Lin,
  Shi, Huang, Lee, and Du}}]{PhysRevB.100.024310}
\bibinfo{author}{\bibfnamefont{T.}~\bibnamefont{Tian}},
  \bibinfo{author}{\bibfnamefont{Y.}~\bibnamefont{Ke}},
  \bibinfo{author}{\bibfnamefont{L.}~\bibnamefont{Zhang}},
  \bibinfo{author}{\bibfnamefont{S.}~\bibnamefont{Lin}},
  \bibinfo{author}{\bibfnamefont{Z.}~\bibnamefont{Shi}},
  \bibinfo{author}{\bibfnamefont{P.}~\bibnamefont{Huang}},
  \bibinfo{author}{\bibfnamefont{C.}~\bibnamefont{Lee}}, \bibnamefont{and}
  \bibinfo{author}{\bibfnamefont{J.}~\bibnamefont{Du}}, \bibinfo{journal}{Phys.
  Rev. B} \textbf{\bibinfo{volume}{100}}, \bibinfo{pages}{024310}
  (\bibinfo{year}{2019}),
  \urlprefix\url{https://link.aps.org/doi/10.1103/PhysRevB.100.024310}.

\bibitem[{\citenamefont{Nie et~al.}(2020)\citenamefont{Nie, Wei, Chen, Zhang,
  Zhao, Qiu, Tian, Ji, Xin, Lu et~al.}}]{PhysRevLett.124.250601}
\bibinfo{author}{\bibfnamefont{X.}~\bibnamefont{Nie}},
  \bibinfo{author}{\bibfnamefont{B.-B.} \bibnamefont{Wei}},
  \bibinfo{author}{\bibfnamefont{X.}~\bibnamefont{Chen}},
  \bibinfo{author}{\bibfnamefont{Z.}~\bibnamefont{Zhang}},
  \bibinfo{author}{\bibfnamefont{X.}~\bibnamefont{Zhao}},
  \bibinfo{author}{\bibfnamefont{C.}~\bibnamefont{Qiu}},
  \bibinfo{author}{\bibfnamefont{Y.}~\bibnamefont{Tian}},
  \bibinfo{author}{\bibfnamefont{Y.}~\bibnamefont{Ji}},
  \bibinfo{author}{\bibfnamefont{T.}~\bibnamefont{Xin}},
  \bibinfo{author}{\bibfnamefont{D.}~\bibnamefont{Lu}}, \bibnamefont{et~al.},
  \bibinfo{journal}{Phys. Rev. Lett.} \textbf{\bibinfo{volume}{124}},
  \bibinfo{pages}{250601} (\bibinfo{year}{2020}),
  \urlprefix\url{https://link.aps.org/doi/10.1103/PhysRevLett.124.250601}.

\bibitem[{\citenamefont{Mei et~al.}(2022)\citenamefont{Mei, Li, Wu, Cai, Wang,
  Yao, Zhou, and Duan}}]{PhysRevLett.128.160504}
\bibinfo{author}{\bibfnamefont{Q.-X.} \bibnamefont{Mei}},
  \bibinfo{author}{\bibfnamefont{B.-W.} \bibnamefont{Li}},
  \bibinfo{author}{\bibfnamefont{Y.-K.} \bibnamefont{Wu}},
  \bibinfo{author}{\bibfnamefont{M.-L.} \bibnamefont{Cai}},
  \bibinfo{author}{\bibfnamefont{Y.}~\bibnamefont{Wang}},
  \bibinfo{author}{\bibfnamefont{L.}~\bibnamefont{Yao}},
  \bibinfo{author}{\bibfnamefont{Z.-C.} \bibnamefont{Zhou}}, \bibnamefont{and}
  \bibinfo{author}{\bibfnamefont{L.-M.} \bibnamefont{Duan}},
  \bibinfo{journal}{Phys. Rev. Lett.} \textbf{\bibinfo{volume}{128}},
  \bibinfo{pages}{160504} (\bibinfo{year}{2022}),
  \urlprefix\url{https://link.aps.org/doi/10.1103/PhysRevLett.128.160504}.

\bibitem[{\citenamefont{Budich and
  Heyl}(2016{\natexlab{b}})}]{PhysRevB.93.085416}
\bibinfo{author}{\bibfnamefont{J.~C.} \bibnamefont{Budich}} \bibnamefont{and}
  \bibinfo{author}{\bibfnamefont{M.}~\bibnamefont{Heyl}},
  \bibinfo{journal}{Phys. Rev. B} \textbf{\bibinfo{volume}{93}},
  \bibinfo{pages}{085416} (\bibinfo{year}{2016}{\natexlab{b}}),
  \urlprefix\url{https://link.aps.org/doi/10.1103/PhysRevB.93.085416}.

\bibitem[{\citenamefont{Porta et~al.}(2020)\citenamefont{Porta, Cavaliere,
  Sassetti, and Ziani}}]{ZianiSR20}
\bibinfo{author}{\bibfnamefont{S.}~\bibnamefont{Porta}},
  \bibinfo{author}{\bibfnamefont{F.}~\bibnamefont{Cavaliere}},
  \bibinfo{author}{\bibfnamefont{M.}~\bibnamefont{Sassetti}}, \bibnamefont{and}
  \bibinfo{author}{\bibfnamefont{N.~T.} \bibnamefont{Ziani}},
  \bibinfo{journal}{Sci. Rep.} \textbf{\bibinfo{volume}{10}},
  \bibinfo{pages}{12766} (\bibinfo{year}{2020}).

\bibitem[{\citenamefont{Nakahara}(2003)}]{Nakahara}
\bibinfo{author}{\bibfnamefont{M.}~\bibnamefont{Nakahara}},
  \emph{\bibinfo{title}{Geometry, Topology and Physics}}
  (\bibinfo{publisher}{Taylor and Francis Group}, \bibinfo{address}{Boca Raton,
  FL}, \bibinfo{year}{2003}), \bibinfo{edition}{2nd} ed.

\bibitem[{\citenamefont{Wang et~al.}(2019{\natexlab{b}})\citenamefont{Wang,
  Qiu, Xiao, Zhan, Bian, Yi, and Xue}}]{WangPRL19}
\bibinfo{author}{\bibfnamefont{K.}~\bibnamefont{Wang}},
  \bibinfo{author}{\bibfnamefont{X.}~\bibnamefont{Qiu}},
  \bibinfo{author}{\bibfnamefont{L.}~\bibnamefont{Xiao}},
  \bibinfo{author}{\bibfnamefont{X.}~\bibnamefont{Zhan}},
  \bibinfo{author}{\bibfnamefont{Z.}~\bibnamefont{Bian}},
  \bibinfo{author}{\bibfnamefont{W.}~\bibnamefont{Yi}}, \bibnamefont{and}
  \bibinfo{author}{\bibfnamefont{P.}~\bibnamefont{Xue}},
  \bibinfo{journal}{Phys. Rev. Lett.} \textbf{\bibinfo{volume}{122}},
  \bibinfo{pages}{020501} (\bibinfo{year}{2019}{\natexlab{b}}).

\bibitem[{\citenamefont{Guo et~al.}(2019{\natexlab{b}})\citenamefont{Guo, Yang,
  Zeng, Peng, Li, Deng, Jin, Chen, Zheng, and Fan}}]{GuoApplied19}
\bibinfo{author}{\bibfnamefont{X.~Y.} \bibnamefont{Guo}},
  \bibinfo{author}{\bibfnamefont{C.}~\bibnamefont{Yang}},
  \bibinfo{author}{\bibfnamefont{Y.}~\bibnamefont{Zeng}},
  \bibinfo{author}{\bibfnamefont{Y.}~\bibnamefont{Peng}},
  \bibinfo{author}{\bibfnamefont{H.~K.} \bibnamefont{Li}},
  \bibinfo{author}{\bibfnamefont{H.}~\bibnamefont{Deng}},
  \bibinfo{author}{\bibfnamefont{Y.~R.} \bibnamefont{Jin}},
  \bibinfo{author}{\bibfnamefont{S.}~\bibnamefont{Chen}},
  \bibinfo{author}{\bibfnamefont{D.}~\bibnamefont{Zheng}}, \bibnamefont{and}
  \bibinfo{author}{\bibfnamefont{H.}~\bibnamefont{Fan}},
  \bibinfo{journal}{Phys. Rev. Applied} \textbf{\bibinfo{volume}{11}},
  \bibinfo{pages}{044080} (\bibinfo{year}{2019}{\natexlab{b}}).

\bibitem[{\citenamefont{Chen et~al.}(2020)\citenamefont{Chen, Hou, Zhou, Qian,
  Shen, and Xu}}]{ChenAPL20}
\bibinfo{author}{\bibfnamefont{B.}~\bibnamefont{Chen}},
  \bibinfo{author}{\bibfnamefont{X.}~\bibnamefont{Hou}},
  \bibinfo{author}{\bibfnamefont{F.}~\bibnamefont{Zhou}},
  \bibinfo{author}{\bibfnamefont{P.}~\bibnamefont{Qian}},
  \bibinfo{author}{\bibfnamefont{H.}~\bibnamefont{Shen}}, \bibnamefont{and}
  \bibinfo{author}{\bibfnamefont{N.}~\bibnamefont{Xu}}, \bibinfo{journal}{Appl.
  Phys. Lett.} \textbf{\bibinfo{volume}{116}}, \bibinfo{pages}{194002}
  (\bibinfo{year}{2020}).

\bibitem[{\citenamefont{Brassard et~al.}(2002)\citenamefont{Brassard, Hoyer,
  Mosca, and Tapp}}]{Brassard02}
\bibinfo{author}{\bibfnamefont{G.}~\bibnamefont{Brassard}},
  \bibinfo{author}{\bibfnamefont{P.}~\bibnamefont{Hoyer}},
  \bibinfo{author}{\bibfnamefont{M.}~\bibnamefont{Mosca}}, \bibnamefont{and}
  \bibinfo{author}{\bibfnamefont{A.}~\bibnamefont{Tapp}}, in
  \emph{\bibinfo{booktitle}{Quantum Computation and Quantum Information}},
  edited by \bibinfo{editor}{\bibfnamefont{S.~J.} \bibnamefont{Lomonaco~Jr.}}
  \bibnamefont{and} \bibinfo{editor}{\bibfnamefont{H.~E.} \bibnamefont{Brandt}}
  (\bibinfo{publisher}{AMS}, \bibinfo{year}{2002}).

\bibitem[{\citenamefont{Dittmann}(1995)}]{REP95}
\bibinfo{author}{\bibfnamefont{J.}~\bibnamefont{Dittmann}},
  \bibinfo{journal}{Rep. Math. Phys.} \textbf{\bibinfo{volume}{36}},
  \bibinfo{pages}{309} (\bibinfo{year}{1995}).

\bibitem[{\citenamefont{Budich and Diehl}(2015{\natexlab{b}})}]{TDMPRB15}
\bibinfo{author}{\bibfnamefont{J.~C.} \bibnamefont{Budich}} \bibnamefont{and}
  \bibinfo{author}{\bibfnamefont{S.}~\bibnamefont{Diehl}},
  \bibinfo{journal}{Phys. Rev. B} \textbf{\bibinfo{volume}{91}},
  \bibinfo{pages}{165140} (\bibinfo{year}{2015}{\natexlab{b}}).

\end{thebibliography}

\end{document}